\documentclass[fleqn,usenatbib]{mnras}

\usepackage[T1]{fontenc}

\DeclareRobustCommand{\VAN}[3]{#2}
\let\VANthebibliography\thebibliography
\def\thebibliography{\DeclareRobustCommand{\VAN}[3]{##3}\VANthebibliography}

\usepackage{graphicx}	
\usepackage{amsmath}	
\usepackage{amssymb}	
\usepackage{hyperref}

\newcommand{\mics}{$\mu$m}
\newcommand{\kms}{km s$^{-1}$}
\newcommand{\ergs}{erg s$^{-1}$}
\newcommand{\jwst}{{\it JWST}}

\title[Dust in the Wind]{GATOS XV: A JWST/MIRI survey of extended circumnuclear dust emission in nearby Seyfert galaxies}

\author[D. J. Rosario et al.]
{David J. Rosario$^{1}$,\thanks{E-mail: david.rosario@ncl.ac.uk}
Houda Haidar$^{1}$,
Steph Campbell$^{1}$,
John Schneider$^{2}$,
Chris Packham$^{2,3}$,
\newauthor 
Nancy A. Levenson$^{4}$,
Almudena Alonso\mbox{-}Herrero$^{5}$,
Richard I. Davies$^{6}$,
Dan Delaney$^{7}$,
\newauthor 
Santiago Garc\'{\i}a\mbox{-}Burillo$^{8}$,
Erin K. S. Hicks$^{7,2}$,
Sebastian F. H\"onig$^{9}$,
Mason Leist$^{2}$,
\newauthor 
Enrique Lopez\mbox{-}Rodriguez$^{10}$,
Miguel Pereira\mbox{-}Santaella$^{11}$,
Anelise Audibert$^{12,13}$,
Enrica Bellocchi$^{14}$,
\newauthor 
Fran\c{c}oise Combes$^{15}$,
Ismael Garc\'{\i}a\mbox{-}Bernete$^{5}$,
Peter Boorman$^{6}$,
Andrew J. Bunker$^{16}$,
Luis Colina$^{5}$, 
\newauthor
Tanio Diaz Santos$^{17,18}$,
Fergus R. Donnan$^{16}$,
Poshak Gandhi$^{9}$,
Omaira Gonz\'alez\mbox{-}Mart\'{\i}n$^{19}$,
\newauthor
Laura Hermosa Mu{\~n}oz$^{3}$,
Alvaro Labiano$^{20}$,
Cristina Ramos Almeida$^{12,13}$,
Claudio Ricci$^{21}$,
\newauthor
Rogemar A. Riffel$^{5,22}$,
Dimitra Rigopoulou$^{16}$,
Daniel Rouan$^{23}$,
Marko Stalevski$^{24, 25}$,
Lulu Zhang$^{2}$
\\\\
The authors' affiliations are listed at the end of the paper.}

\date{Accepted XXX. Received YYY; in original form ZZZ}
\pubyear{2023}
\begin{document}
\label{firstpage}
\pagerange{\pageref{firstpage}--\pageref{lastpage}}
\maketitle

\begin{abstract}

The subarcsecond angular resolution and stable background of  \jwst\ has given us the first high-fidelity images of the arcsecond-scale environment around Active Galactic Nuclei (AGNs) in the nearby Universe. With mid-infrared (MIR) surface brightness sensitivities that are much deeper than the best ground-based instruments, the Mid-InfraRed Instrument imager (MIRIM) now allows us to understand the structure and thermal properties of dust using information over wavelengths of $5$--$25$ \mics, almost all of the MIR range. We present a Cycle 1 \jwst\ MIRIM survey of Seyfert galaxies with the express aim of characterising AGN-heated dust in the central few 100 pcs, and searching for signatures of dust-laden nuclear outflows. This paper outlines the motivation behind the programme, the data reduction and analysis techniques used to isolate the nuclear and extended emission, and a comparison of the observed MIR structures with those seen in other phases (stars, ionised and molecular gas, absorbing dust). In concert with earlier studies that used these data, we conclude that resolved AGN-heated dust is widespread in the Seyfert population, extending out to a few hundred pcs from the nucleus and often displaying a higher surface-brightness compared to the more widespread star-forming dusty circumnuclear disk. Even after accounting for contamination from emission lines in the MIRI filters, we find strong spatial correlations between MIR dust emission and the AGN-ionised gas in the narrow-line region (NLR). 

\end{abstract}




\section{Introduction}

AGN are essential components of the ecosystem of massive galaxies, through the processes of gas accretion from the ISM and feedback of energy back into galaxies. In this cycle, interstellar dust is a crucial component, its unique properties playing an important role in the connections between supermassive black holes and galaxies \citep{alexander12, ramosalmeida17,  harrison18, harrison24, alexander25}.

Dust in the interstellar medium (ISM) is composed primarily of amorphous or partially-crystalline grains of graphite or silicates, with a distribution of sizes that range from nanometres to microns (larger grains will exist, but have effects that are difficult to observe). In addition, large complex organic molecules, such as polycyclic aromatic hydrocarbons (PAHs), are also widespread in the ISM, and are often considered as components of dust. For a detailed discussion of the phenomenology and physics of interstellar dust, we refer the reader to the classic review of \citet{draine03}, as well primary work on ISM dust models \citep[e.g.][]{draine07, hensley23}.

Dust is the prime determinant of obscuration at ultraviolet--optical (UVO) wavelengths, and largely governs our observed view of AGN in this regime. In this fashion, dust is the key component of the ``obscuring torus'', the dense concentration of material that is shown to reside within the inner few parsecs all AGN with radiatively-efficient accretion disks. This dust is responsible for the apparent dichotomy of AGN types in the UVO: optical Type\,1 AGN have minimal dust obscuration along the line-of-sight, allowing a clear view of the central accretion disk, while optical Type\,2 AGN are subject to at least a few magnitudes of dust extinction towards the nucleus and only exhibit emission lines produced in the narrow-line region (NLR; material at $\gtrsim 100$ pc from the nucleus influenced by the radiation field from the AGN).

Related to its powerful capacity to absorb UVO light, dust is also known to mediate the coupling between the radiation from an AGN and the ISM. The optical depth to UVO radiation from dust can be an order of magnitude larger than that from gaseous or plasma scattering. This implies that the primary process by which radiation interacts with the dusty, dense ISM (such as in the AGN torus) is through the dynamical response of dust, which in turn, transfers its acquired momentum to surrounding gas \citep[e.g.,][]{ishibashi15}.

Studies of the relation between nuclear obscuration and AGN luminosity have identified the important role that dust-mediated radiation pressure plays in the structure and time-dependent nature of the torus \citep[e.g.,][]{ricci17, lansbury20,drewes25,ramosalmeida26}. An important theoretical prediction of the strong coupling between AGN radiation and the dust of the torus is the presence of radiation pressure-driven dusty winds. These winds are launched primarily at the inner edge of the torus, around the radius at which dust sublimes at temperatures of $T \approx 1000$--$1500$ K. In these regions of intense interaction between the radiation field of the AGN and dust-laden inflowing gas, small and silicate-rich grains are expected to be preferentially destroyed, while larger graphitic grains are carried aloft in a wind that, guided by pressure gradients from the torus itself, moves primarily along an axis that is perpendicular to the AGN disk \citep[the eponymous ``polar dusty wind''][]{hoenig19}.

In radiation-hydrodynamic simulations of AGN tori, polar winds are a common feature. Indeed, such winds have been shown to travel a large distance from the nucleus, resulting in a polar extended dusty structure which can be as much as 200 pc in size \citep{williamson19,williamson20}. With temperatures of 100s of K, these winds are expected to emit primarily in the mid-infrared (MIR) regime for most disk+wind configurations \citep{alonso-herrero21}


If polar dusty winds are ubiquitous in Seyfert AGN, and can propagate to scales of 100s of pc, they may be responsible for the tentative reports of polar-aligned MIR emission seen in ground-based images of a modest fraction of local AGN \citep{asmus16, asmus19, stalevski19, alonso-herrero21}. Despite the unprecedented spatial resolution of the MIR imagers on ground-based 8-10 m class telescopes, the strongly variable nature of the bright atmospheric and thermal foreground has always limited the sensitivity of these early studies, particularly to the low-surface brightness emission that dusty extended winds may exhibit.

With the advent of JWST, the largest space telescope ever launched, we are now able to detect emission in the sub-arcsecond vicinity of AGN nuclei  down to surface brightness levels orders of magnitude fainter than ground-based instruments in the thermal infrared. In Cycle 1 of JWST observations, we conducted an imaging programme (ID 2064: `Dust in the Wind') with the Mid-Infrared Instrument (MIRI), designed to explore the central emission of a sample of nearby AGN with robust evidence for extended MIR structure from earlier studies. Our aim was to understand the diversity of the dust emission, its relationship to other gaseous phases (ionised and molecular gas) and, ultimately, search for evidence of radiatively-accelerated dusty outflowing gas on circumnuclear scales of $\sim 100$ pc. 

This reference publication is the main documentation for the motivation and approach of the PID 2064 programme, as well as the techniques we have employed in putting together the first science results based on these observations \citep{haidar24, campbell25, haidar26}.

The other \jwst\ instrument mode that is widely used for nearby AGN science is the MIRI Medium-Resolution Spectrograph (MRS), an integral-field unit (IFU) that with a field of $\approx 4$--$7$ arcseconds. The MRS has sustained considerable work on emission lines and polycyclic aromatic hydrocarbons (PAHs), with a range of fascinating astrophysical insights \citep[e.g.,][]{davies24, garcia-bernete24b, hermosamunoz24, zhang24, esparza25, ramosalmeida25}. Only a few MRS studies have explored the resolved dust continuum, mainly due to the intricacies introduced by the complex instrumental point-spread function (PSF) over the small MRS field. A pioneering MRS-based study from \citet{lopez25} explored the nature of the dust continuum and its spatial relationship with MIR emission lines at similar wavelengths (e.g., [Fe VII]$\lambda 7.82$~\mics, [S IV]$\lambda 10.51$~\mics, [Ne V]$\lambda 14.32$~\mics, [S III]$\lambda 18.71$~\mics). They reported a fairly low level of correlation, implying that the warm dust emission is not strongly connected to the line-emitting regions, which, in AGN, are dominated by the NLR.

The MIRI imager has some advantages for continuum work over the MRS: better surface-brightness sensitivity, as well an angular resolution that is 25\% narrower and better modelled than the MRS at the comparable wavelengths. In recent work based on PID 2064, \citet{haidar26} reported strong relationships between the dust continuum structure and high-ionisation line emission from the AGN. We will examined these apparent differences in more detail in Section \ref{spatial_correlations}.

Unless otherwise specified, all magnitudes and colours are on the AB system. All the targets are within 66 Mpc, so we take a locally-Euclidean approximation to cosmological parameters, such that luminosity distances ($D_{L}$) are equal to angular-diameter distances. For the furthest target (NGC\,5135; $D_{L} = 64.8$ Mpc), this assumption is correct to better than 3\%. Redshift-independent distance estimates have been used based on the NED-D compilation \citep{steer17}.

\section{Observations and Datasets}

\subsection{JWST observations}

\begin{figure}
	\centering
	\includegraphics[width=\columnwidth]{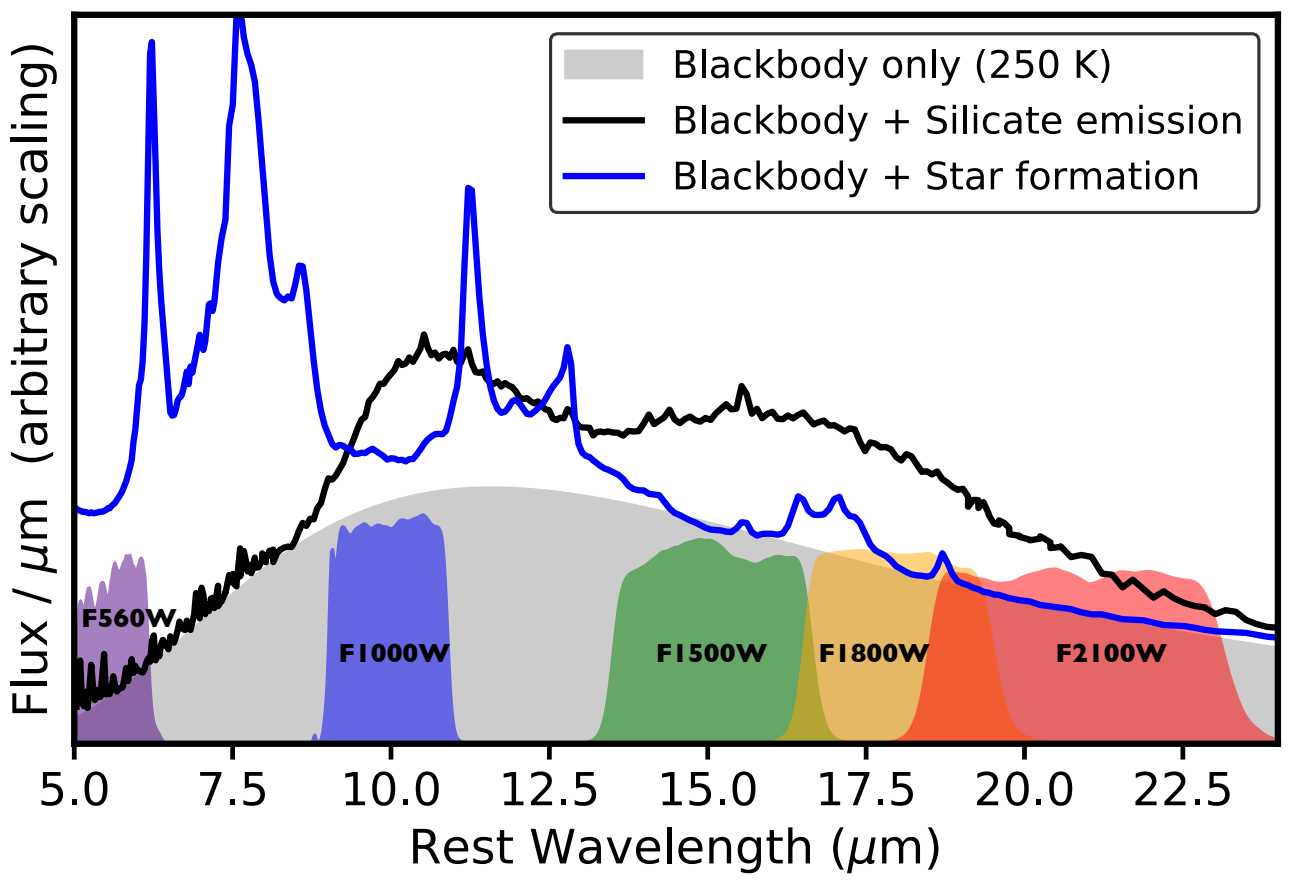}
	\caption{A schematic that demonstrates the choice of filters used in the MIRIM imaging programme. The black solid line show a characteristic AGN-like SED consisting of excited silicate emission over dust continuum of a fixed temperature of 250 K. The blue solid line shows dust of a similar temperature, but overlaid by the bright PAH features typical of star-forming regions. The filters sample as much of the SED as possible, while avoiding the bright PAH bands between 5 and 13 \mics\ which are difficult to model.}
	\label{filter_choice}
\end{figure}

\begin{table*}
	\caption{Details of the targets of the PID 2064 MIRI imaging programme}
	\begin{minipage}{\textwidth}
	\begin{center}
	\begin{tabular}{lcclc}
	\hline
	\hline
	Galaxy Name & Nuclear RA, DEC (J2000, deg)\footnote{Coordinate references: NGC\,4388: \citet{kamali17}, NGC\,3227: \citet{alonso19}, ESO 428-G14: \citet{falcke98}, NGC\,3081: ALMA continuum, NGC\,7172: \citet{alonso23}, NGC\,2992: \citet{fernandez22}, NGC\,5728: \citet{shimizu19}, NGC\,5135: NICMOS image} & Distance (Mpc)\footnote{From the NASA Extragalactic Database, using, where available, averages of redshift-independent distances published after 2000.} &  AGN Type\footnote{Seyfert Type based on optical spectroscopy. Sy2i$=$infrared broad lines. Sy2h$=$broad lines in polarised light} & MIRIM Observation date \\
	\hline
	NGC\,4388 & 186.444917, 12.66216 & 18.1 & \hspace{0.11in}Sy 2h & 27 Jun 2023  \\
	NGC\,3227 & 155.877391, 19.86508 & 23.0 & \hspace{0.11in}Sy 1.5 & 26 Nov 2022  \\				
	ESO 428-G14 & 109.130029, -29.32469 & 23.2 & \hspace{0.11in}Sy 2 & 21 Oct 2022 \\
	NGC\,3081 & 149.873110, -22.82633 & 32.5 & \hspace{0.11in}Sy 2h & 31 May 2023 \\
	NGC\,7172 & 330.507875, -31.86958 & 33.9 & \hspace{0.11in}Sy 2 & 04 Nov 2022  \\		
	NGC\,2992 & 146.424760, -14.32627 & 38.0 & \hspace{0.11in}Sy 2i & 27 Apr 2023 \\								
	NGC\,5728 & 220.599466, -17.25306 & 39.0  & \hspace{0.11in}Sy 2 & 03 Mar 2023  \\
	NGC\,5135 & 201.433278, -29.83334 & 64.8 & \hspace{0.11in}Sy 2 & 02 Mar 2023  \\
	\hline
	\hline
	\end{tabular}
	\end{center}
	\end{minipage}
	\label{obs_details}
\end{table*}

Eight Seyfert galaxies were targetted by PID 2064, all with evidence for extended MIR emission at $\approx 10$ \mics\ from ground-based instruments on 8-10 m class telescopes \citep{asmus19}. Some very nearby AGN, such as NGC\,1068 and Circinus, were avoided because their nuclei are too bright for even the shortest available JWST observations. Besides this, no specific lower flux limit was placed on our selection. In the spirit of a pilot study, the targets were chosen to cover a range of AGN luminosities and Eddington ratios, and all have a wealth of high-resolution ancillary data (see Section \ref{mwcomps}). Details of our targets are listed in Table \ref{obs_details}, sorted in increasing order of distance. The \jwst\ dataset can be accessed from the Multi-mission Archive for Space Telescopes (MAST) at \href{https://doi.org/10.17909/yv20-cm46}{doi.org/10.17909/yv20-cm46.}

The observations with the \jwst/MIRI imager (MIRIM, henceforth) were taken over the course of Cycle 1. Observing sequences for a single target were split into a pair of `visits'.
An on-source visit consisted of 5 sets of exposures, each taken through one of five MIRIM filters: F560W, F1000W, F1500W, F1800W, F2100W. 
The reasoning behind the choice of MIRI filters can be understood from Figure \ref{filter_choice}.
The filters were selected to cover as much of the continuum MIR emission from the central nucleus and surrounding
circumnuclear star-forming regions, while avoiding strong contamination from PAH emission. Various filters are
sensitive to emission from the smooth underlying dust continuum, and emission/absorption from silicate features at 9.7 \mics\ and 18 \mics.

A 4-point dither sequence was
employed for each set of on-source exposures. An important part of the science case requires an optimal reconstruction of the bright point source at the nucleus of all our targets (Section \ref{scireduction}). Therefore, we utilised a filter-specific point source-optimised dither sequence.

Each on-source visit was immediately followed by a background visit to allow optimal removal of the thermal
and zodiacal backgrounds which affect \jwst\ MIR observations. 
The background visits employed the same series of filters as the on-source visits, and their pointings were 
placed on source-free regions of sky typically a few arcminutes away from the target
galaxies. This ensured that the outskirts of the galaxies did not significantly contaminate the background emission.
A 2-point dither was used for the background exposure sequences, to ensure redundancy and deal with cosmic
rays and bad pixels. 

A MIRIM subarray (SUB256; $\approx$ 28" on a side) was used for the observations, which enabled a frame readout time ($0.3$ seconds) that was short enough to avoid saturation of the bright nuclear emission from the AGN. Each exposure consisted of 50 integrations of 5 groups up-the-ramp, for a total effective on-source time of 358 seconds after combining the 4 dithered exposures. The ramp lengths were the shortest recommended by pre-flight documentation for \jwst\, and were chosen to minimise the number of saturated pixels in the images. The total integration times were designed to detect the known arcsecond-scale extended MIR emission in our targets with signal-to-noise ratios $> 10$ per pixel, based on pre-flight \jwst\ exposure time calculations.

The first observations of three galaxies were mispointed, placing the targets near the field edges for some of the frames. Here we report on subsequent reobservations that had the correct pointing. See Appendix \ref{firstobs} for more details.

\subsubsection{JWST data reduction} \label{scireduction}

\begin{figure*}
	\centering
	\includegraphics[width=\textwidth]{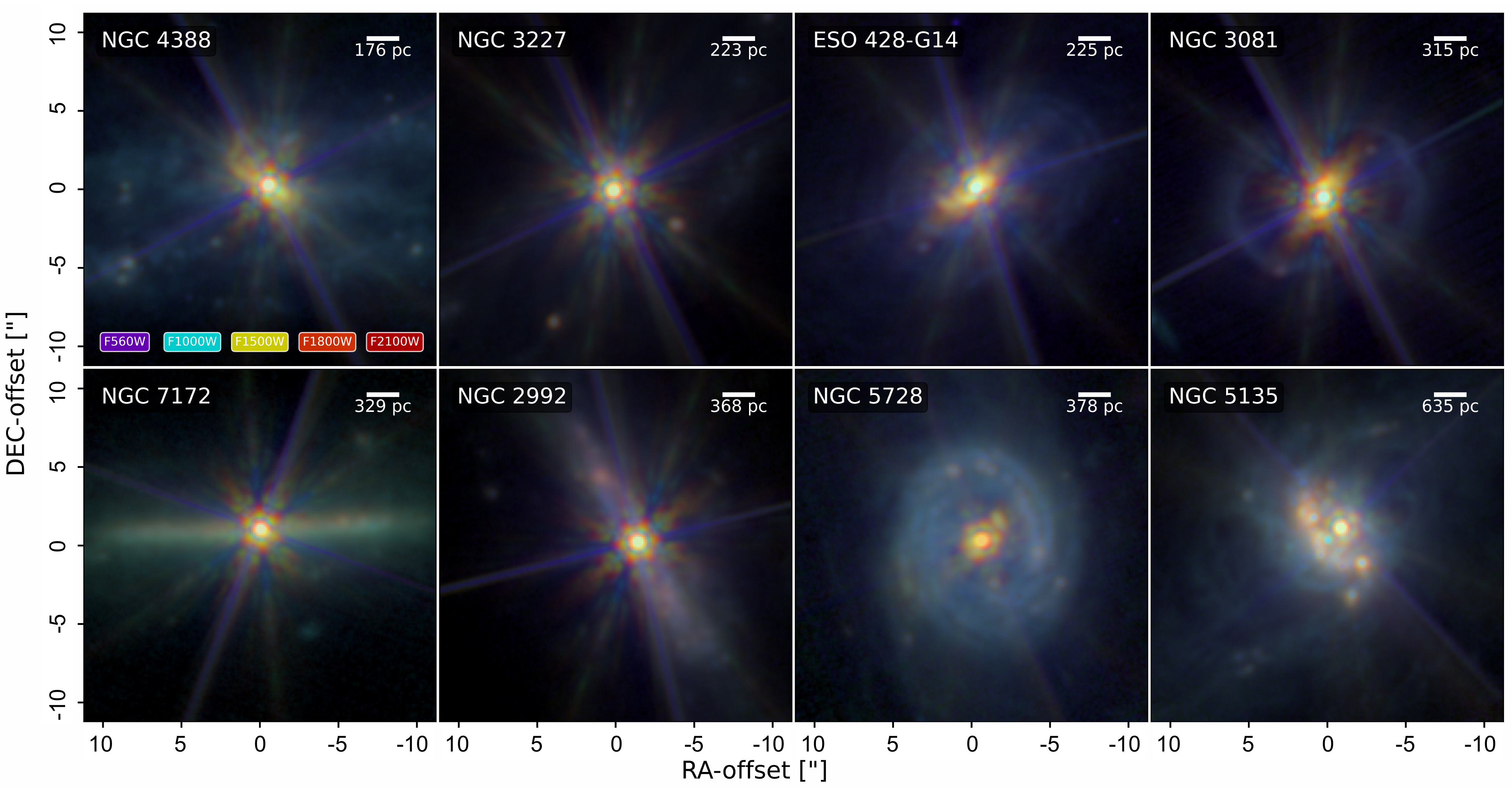}
	\caption{False colour images of all eight Seyfert galaxies targetted by our JWST imaging programme, created by mapping the fluxes from each filter to a particular colour as shown in the key. All images are scaled and stretched in the same way to allow a quick comparison of the colours of the various visible components. The images are oriented with North to the top and East to the left. The bar denoting 2'' is labelled with the equivalent projected physical scale at the adopted distances of the targets. Bright nuclear point sources are visible in all images, even though seven of our targets are optical Type\,2 AGN.}
	\label{all_RGB}
\end{figure*}

Bright MIR nuclear emission is expected in all of our targets, because of their nearness and
modest nuclear luminosities. This sets up a particular challenge for our observing and data reduction strategy: 
we require images with a large dynamic range so that faint circumnuclear emission can be uncovered after optimal
removal of the core and wings of the central point source.

The standard \jwst/MIRI data calibration pipeline \citep{jwstpipe} is multifarious, but its default configuration does not properly
handle the specific features of our images. Therefore, we altered various steps of the reduction pipeline and
added some alternate specialised reduction steps to produce the final science-grade images used in the rest of this work. All reduction was performed with version 1.19.2 of the \jwst\ calibration pipeline, with Calibration Reference Data System (CRDS) version 13.0.6 and CRDS context \texttt{jwst\_1413.pmap}.

We downloaded the raw uncalibrated data from MAST. 
The primary science data are in the form of UNCAL files which store full images obtained at every up-the-ramp multiaccumulation read of the detector for each group and integration in an exposure. The detector-level reduction of the UNCAL files is performed by Level 1 of the MIRIM pipeline (\texttt{calwebb\_detector1}). Most relevant to our programme, \texttt{calwebb\_detector1} checks for saturated groups, detects jumps within integrations, and fits the ramps to calculate linearised electron fluxes per pixel for each integration and exposure.
In pixels with partial or total saturation, the ramps deviate strongly from a linear time series. The pipeline identifies these deviant timelines, excludes fully saturated groups and uses a correction to extrapolate linearised rates from the remaining groups. 

In ramps with strong saturation, only the first and second groups may be measurable, with later groups flagged as partially or completely saturated. This limitation is exacerbated by the default behaviour of the pipeline, in which the first group in a ramp is suppressed (``first frame correction'') because the detector reset performed just prior to a new ramp can produce a spurious signal transient in the first read. After the default flagging of the first group, many ramps in regions of high surface brightness are left with only a single unflagged group measurement, and are therefore marked as ``saturated''. Because the brightness of the source emission on these pixels is very high, the detector reset transient is only a small source of noise. Therefore, after consulting with the MIRIM imaging team, we turned off ``first frame correction'' for pixels that were identified as ``saturated'' in the default pipeline output of \texttt{calwebb\_detector1}. This intervention effectively removed any saturated pixels in all the exposures.

We visually examined the light profiles of the points sources in all targets after these pipeline interventions. In all cases, we found excellent agreement with the profiles of model PSFs for the respective filters (Section \ref{psfs}). 

Level 2 pipeline reduction (\texttt{calwebb\_image2}) initialises the world coordinate system (WCS) keywords in the images, performs background removal using offset frames, and applies flat-field and flux calibrations. As none of these steps are affected by the nature of our targets, we used default parameters in the pipeline for this stage.

The outputs of \texttt{calwebb\_image2} are four on-target images in each filter band (exposures, henceforth), corresponding to the four dither positions used during the observations.  

The Level 3 stage of the \jwst\ pipeline (\texttt{calwebb\_image3}) produces science-ready images. The first task of this stage tweaks the astrometry of
the Level 2 outputs to achieve sub-pixel registration, a
necessary step towards the creation of final combined, distortion-corrected image combinations from the dithered exposures. The default algorithm for tweaking the astrometry requires the detection of multiple compact sources (e.g., stars or distant galaxies) from the images.
However, this approach is not suitable for the observations of our targets, since the large host galaxy fully fills the field-of-view of the detector subarray. The short exposure times and long thermal MIR wavelengths of our images also limits the detectability of faint stars or background galaxies.
Instead, we manually assessed the relative astrometry of the four dithered exposures using the bright nuclear point source of the AGN as a positional reference. 

For each exposure, we iteratively determined the centroids of the nuclear emission in 5'' boxes around the nucleus, using a two-dimensional Gaussian centroiding algorithm from the PHOTUTILS Python package. We then compared the relative astrometry of the centroids of the images from each exposure, and visually assessed the alignment using difference images generated from pairs of exposures in the same filter band. In all the dither sequences taken over the many observations for the programme, the individual exposures were always registered to better than 1/3 of a native pixel (37 mas). Therefore, we were able to combine the subexposures without any astrometric tweaks. 



We processed the exposures with the rest of the steps of the Level 3 pipeline. 
The final step of the Level 3 pipeline combines the exposures with the Astrodrizzle algorithm to produce final distortion-corrected images. We used the native MIRIM pixel scale ($0\farcs11$) for all the images except for those taken with the F560W filter. At these shorter wavelengths, the MIRIM detector undersamples the PSF. To take full advantage of our point source-optimised dithering strategy, we produced final F560W images with a pixel scale of $0\farcs07$, slightly higher than the Nyquist sampling scale for the PSF FWHM.  

Finally, we placed our images onto an absolute astrometric frame by calculating the coordinate shift-offset needed to place the centroid of the nuclear point source from the highest resolution images (oversampled F560W) at the absolute nuclear position of each target 
(see Table \ref{obs_details}). We assumed that the relative astrometry is accurate between the Level 3 images 
in all the filter bands. Since there were no guiding issues during the on-source visits for any of our targets,
this assumption is reasonable. 

The ultimate science-ready dataset from our adapted pipeline reduction consists of five images for each target, one for each of the five filter bands, with associated error and weight maps, and relevant metadata.

Figure \ref{all_RGB} shows false-colour visualisations of our targets generated from the processed and astrometrically-aligned final images. These images were constructed by a weighted mapping of the wavelengths of each JWST filter into the visible RGB channels, along with logarithmic stretching and white-balancing to enhance the visibility of extended emission\footnote{See \citet{schneider26} for details}. While all galaxies show dominant central emission from the MIR-bright nucleus, a range of resolved structure is also seen, such as circumnuclear disks and spirals, bipolar features, and compact off-nuclear knots of emission.

\subsection{Point-spread functions (PSFs)} \label{psfs}

An important prerequisite to the data presentation and analysis in this work is the modelling and 
subtraction of the nuclear point source from the MIRIM images. This requires the accurate characterization and modelling of the imaging PSFs. 
For this, we used STPSF (formerly WebbPSF), a
Python-based package to generate PSF models for \jwst\ observing modes \citep{perrin12}, maintained by STScI to support \jwst\ science.

Even at optimal focus, the PSF of \jwst/MIRI varies strongly as a function of wavelength and, 
more subtly, across the field of the detector. The PSF also varies over time
due to changes in the focus and optical wavefront, thermal variations in the instrument, and rapid optical and mechanical jitters across the whole system. 
WebbPSF uses prescriptions for the optical path differences (OPDs) introduced by the telescope and the instrument to calculate PSFs across the field, for various filters and input source spectral energy distributions. It also produces time-dependent PSFs using OPD information derived from calibrations taken closest in time to a user-specified date.\footnote{
More details can be found on the WebbPSF documentation pages: \href{https://webbpsf.readthedocs.io}{webbpsf.readthedocs.io}}.

In parallel with the Level 3 pipeline processing of the science images (Section \ref{scireduction}), we generated simulations of dithered and drizzled PSF images at the location of the nuclear point source on the imager field. For each filter band, we first produced model PSFs at the target's nuclear pixel positions for each exposure, sampled at the native detector resolution and including the distortion of the detector field. We then ran these simulated PSF exposures through the modified Level 3 pipeline with the same drizzling and de-distortion parameters that
were used for the combination of the science images. This
yielded a set of five model PSF images per target that faithfully
captured the instrumental and the detector-sampling
elements of the PSFs, at the correct location of the nucleus in the respective science images. 



\subsection{Subtraction of the nuclear ``point source''} \label{psf_subtraction}

\begin{figure*}
	\centering
	\includegraphics[width=\textwidth]{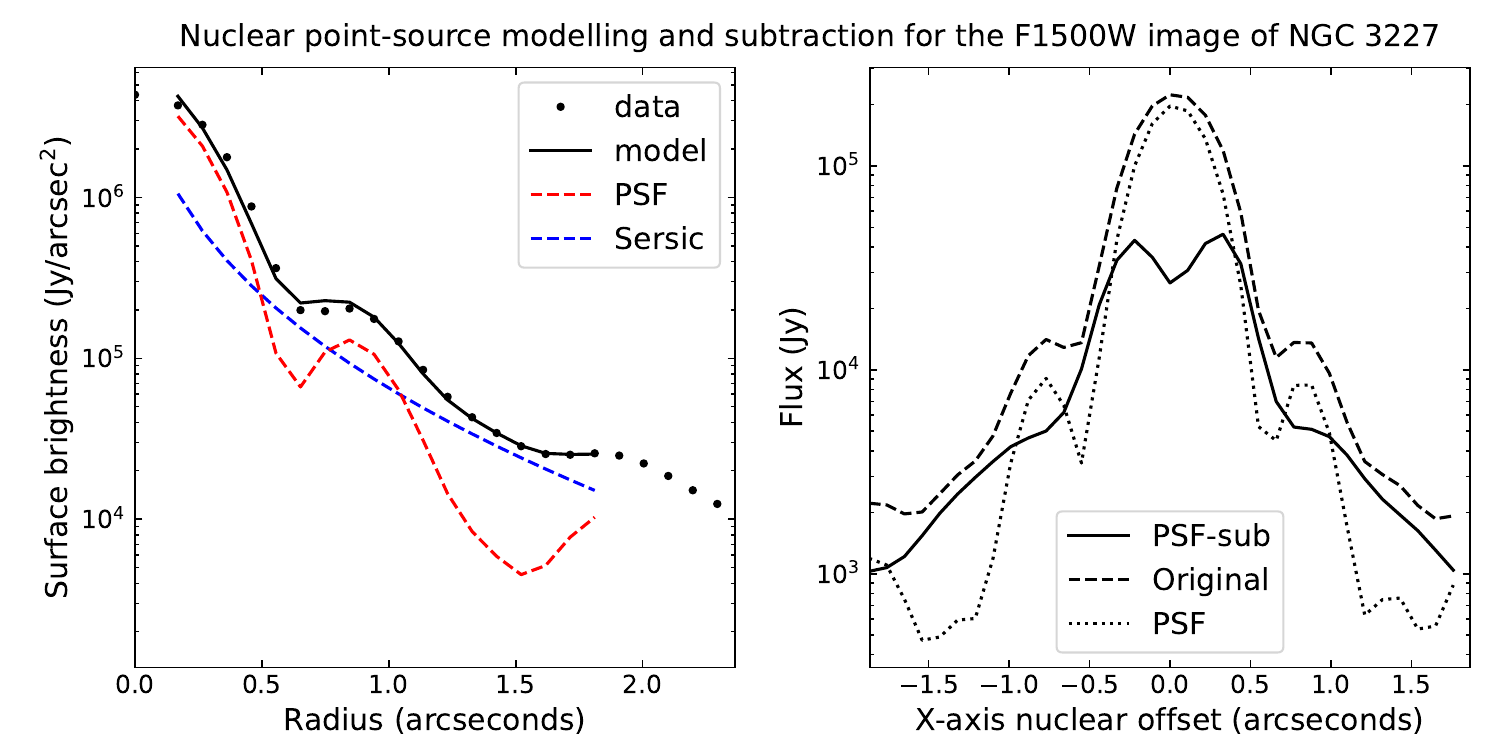}
	\caption{An example of the fit of the nuclear point source in the F1500W image of NGC\,3227, the target with the most prominent nuclear emission in the sample. The left panel shows the azimuthally-averaged light profile measured from the image centred on the nucleus (black dots). This profile is fit to a model (black line) which can be decomposed into components from the PSF and underlying extended galaxy light (red and blue dashed lines respectively). Note that the fit is primarily constrained by the contrast of the bumps in the PSF light profile which comes from Airy diffraction features. The right panel shows an extracted profile from a collapsed 5 row-wide region of the original image through the nucleus (dashed line) compared with same extraction from a PSF-subtracted image (solid line).	
	}
	\label{psf_modelling_example}
\end{figure*}

The stability of the \jwst\ PSF allows us an unprecedented view
of the near-nuclear dust environment in our targets. However, the PSF itself displays considerable structure, such as a bright core, a hexagonal anisotropic Airy ring, a set of six diffraction spikes, and a 4-armed cruciform pattern that arises from scattering within the detector \citep{gaspar21}. Therefore, to bring out the structure in the immediate region of the nucleus, we developed
an approach to model and subtract a nuclear point-source from the images. 

We assumed that the light distribution around the nucleus consists of a bright point-source superimposed over a smoother profile from underlying partially-resolved and fully resolved emission.
To first order, the assumption of a nuclear point source is reasonable because the majority of the nuclear MIR emission in the AGN should come from the inner torus which, at scales of a few parsecs, is completely unresolved in our images. 

From Figure \ref{all_RGB}, it is clear that the extended emission over the whole field of our MIRIM images is not smooth or uniform. To alleviate the effects of circumnuclear substructure, we azimuthally-averaged both the image and the PSF light profiles before undertaking any further modelling.
While the PSF is not circular, it does show an azimuthal regularity imposed by the hexagonal shape of the primary mirror and its segments. The circumnuclear structures of the host galaxies themselves are not as regular. Therefore, we adopt a S\'ersic light profile for the extended light to allow for a fair degree of freedom in the modelling of this component. In practice, because of the dominance of the PSF structure from the central point source, the fits are primarily constrained by the contrast of the Airy ring and diffraction spikes as they appear in the azimuthally-averaged light profile (see Figure \ref{psf_modelling_example}). The S\'ersic function simply serves as a handy centrally-peaked smooth model for the underlying extended emission.

The one-dimensional S\'ersic function for the underlying galaxy has three free 
parameters: the S\'ersic index ($n$), the effective radius ($r_{e}$), and the normalisation. We fixed the centre of the
profile at the known position of the nucleus, which is
accurate to better than $0\farcs07$. We used the model PSFs
(see Section \ref{psfs}) to calculate the expected surface-brightness profile of the nuclear point source, also centred on the nuclear position.
Using a non-linear least-squares fitting algorithm, implemented through CURVEFIT in SciPy \citep{scipy}, we fit for the
scaling of the nuclear point-source while 
allowing the parameters that described the galaxy to vary
within reasonable bounds: strictly positive 
fluxes for the point-source and galaxy, 
$r_{e} < 20$'', $n<8$. We restrict the fit to radii $<3 \times$ the 80\% encircled-energy radius (EER80) of the PSF, which is sufficient to track the main PSF features which constrain the fit.

An example of the outputs from this fitting procedure is shown in Figure \ref{psf_modelling_example}. The left panel shows the 1-dimensional light profile of NGC\,3227 in the F1500W image, from the nucleus to a radius of 3 arcseconds. The EER80 of the F1500W MIRIM PSF is $0\farcs93$; $3 \times$ this radius sets the maximal nuclear distance to which we perform the fit, which is indicated in the figure by the radius at which the model lines cut off.

We used the point-source normalisation from our combined fit to scale the PSF model and subtract it from the image to yield a point source-subtracted residual image. 

It is valuable to examine the performance of the point source subtraction to understand some of the features of the residual images used in later analysis. In the right panel of Figure \ref{psf_modelling_example}, we plot a comparison of the original and point source-subtracted central light profile of the F1500W image of NGC\,3227, the target with the strongest nuclear point-source contrast. The point-source model removes the azimuthally-averaged first and second Airy features quite well, while also reducing the core brightness by $\approx 5$. In the very core of the residual, we often see some oversubtraction around the nucleus which takes the form of a dip in the light profile at the nucleus. This is more pronounced in the images with a brighter point source, and likely indicates a remaining effect of the non-linearity of the detector in partially-saturated ramps (see Section \ref{scireduction}). For this reason, we avoid using any pixels within the inner half of the PSF FWHM for the light profile modelling described above.

Another feature of the residual light profile are the ripples at radii just beyond the first and second Airy bumps of the PSF. These are the corresponding Airy features produced by the bright extended emission underlying the nuclear point-source, which are not modelled in our current approach. In the point source-subtracted images shown in subsequent figures in this work, weak diffraction patterns are the two-dimensional manifestations of the ripples described above. They can only be minimised through deconvolution analysis \citep[e.g.,][]{leist24}, which is beyond the scope of this work.

In the spirit of an overview paper, we proceed with our current modelling approach while acknowledging the limitations of the method.
We have remained cognizant
of the presence of remaining diffraction residuals in Section \ref{results}, as well as other papers based on this MIRIM survey \citep{haidar24, campbell25, haidar26}.

\subsection{VLT/SINFONI emission line maps} \label{si6}

We employ a set of maps of the [Si VI]$\lambda 1.96$ \mics\ emission line from the LUNIS-AGN project, based on K-band observations with the Spectrograph for INtegral Field Observations in the Near-Infrared \citep[SINFONI;][]{eisenhauer03,bonnet04} instrument on the Very Large Telescope (VLT). Details of the raw datasets, reduction, processing, and map creation are in the LUNIS-AGN survey paper \citep{delaney25}. The native absolute astrometry of SINFONI is very uncertain, so we have tied the nuclear positions of our targets to the centroids of the K-band continuum images associated with the [Si VI] maps.

Except for NGC\,3227 and NGC\,5135, the SINFONI observations for the galaxies were Adaptive-Optics assisted, yielding typical angular resolutions of $0\farcs17$, very comparable to those of \jwst\ in the MIR. The seeing-limited SINFONI observations of NGC\,5135 have a PSF FWHM of $0\farcs62$, and those of NGC\,3227 have a FWHM of $0\farcs55$.

The [Si VI]$\lambda 1.96$ \mics\ line is only produced by highly excited gas under the influence of an AGN's radiation field or very fast shocks. In this work, we use these line maps as spatial templates of the NLRs to compare to the structures we see in the \jwst\ images.

\subsection{{\it Hubble Space Telescope} imaging}

\begin{table*}
	\caption{Details of the HST imaging datasets}
	\begin{minipage}{\textwidth}
	\begin{center}
	\begin{tabular}{lcccccc}
		\hline
		\hline
		Galaxy Name & Optical Imager & Optical Filter & Optical Prog. ID & NIR Imager & NIR Filter & NIR Prog. ID \\
		\hline
		NGC\,4388 & WFC3-UVIS & F438W & 12185 & WFC3-IR & F160W & 12185 \\
		NGC\,3227 & WFC3-UVIS & F547M & 11661 & NICMOS & F190N & 11080 \\
		ESO 428-G14 & WFPC2 & F814W & 5411 & --  & -- & --\footnote{No NIR imaging in the {\it HST} archive}\\
		NGC\,3081 & WFPC2 & F606W & 5479 & NICMOS & F160W & 7330 \\
		NGC\,7172 & WFC3-UVIS & F606W & 15181 & WFC3-IR & F160W & 15181 \\	
		NGC\,2992 & WFPC2 & F606W & 5479 & NICMOS & F200N & 7869 \\
		NGC\,5728 & WFC3-UVIS & F438W & 13755 & WFC3-IR & F160W & 13755\\				
		NGC\,5135 & WFPC2 & F606W & 5479 & NICMOS & F160W & 10169 \\
		\hline
		\hline								
	\end{tabular}
	\end{center}
	\end{minipage}
	\label{hst_data}
\end{table*}

Where available on MAST, we obtained {\it Hubble Space Telescope (HST)} optical and near-infrared (NIR) broadband images for our targets. The images were taken by a variety of instruments on {\it HST}: the Wide-field and Planetary Camera 2 (WFPC2), the Near-infared Camera and Multi-object Spectrometer (NICMOS) and the Wide-field Camera 3 (WFC3). Details of the corresponding observations are given in Table \ref{hst_data}.

For all targets except for NGC\,5135, we used versions of the reduced images from the Hubble Legacy Archive (HLA). These are processed by the default instrument pipelines, at a level that is suitable for the heuristic assessment used in this work. The absolute astrometry is inaccurate at the level of $\sim 1$ arcsecond for HLA outputs, though the relative astrometry across the images is good to the subpixel level. In order to allow a proper comparison to the {\it JWST} and other images, we corrected the astrometry of the {\it HST} optical images using the known GAIA positions of stars identified in the images. We chose 1--3 unsaturated stars with measured GAIA coordinates that lay within 30 arcseconds of the centres of our target galaxies, and worked out the average offsets in RA and DEC that allowed the coordinates of the stars to be placed on their centroidal positions in the images. Rotational corrections were not used because they are difficult to estimate with only a few astrometric reference stars and would only have minor effects for our current purposes.

The small field-of-view of the NICMOS images prevented the identification of alignment stars in some cases. For these images, we tied the centroidal position of the galaxy itself to the known nuclear position of the AGN (Table \ref{obs_details}). Underlying this approach is the assumption that either the stellar emission of the host galaxy's bulge, or alternatively the nuclear emission of the AGN, is sufficiently unobscured by dust in the NIR that the centre of light in these images is an accurate marker of the nucleus. 

Processed and science-ready images of NGC\,5135 were first presented in \citet{colina12}, and subsequently astrometrically-aligned to the GAIA reference frame by co-authors of this work.  

\subsection{ALMA CO maps}

\begin{table*}
	\caption{Details of the ALMA observations}
	\begin{tabular}{lcccc}
		\hline
		\hline
		Galaxy Name & ALMA Prog. ID & CO transition & Beam FWHM (mas) & Data source \\
		\hline
		NGC\,4388 &  2017.1.00082.S  & 3-2 & $131 \times 127$ & \citet{garcia21} \\
		NGC\,3227 &  2016.1.00254.S  & 2-1 & $214 \times 161$ & \citet{alonso19} \\
		ESO 428-G14 &  2015.1.00086.S  & 2-1 & $679 \times 586$ & ALMA Archive\\
		NGC\,3081 &  2015.1.00086.S  & 2-1 & $574 \times 480$ & ALMA Archive  \\
		NGC\,7172 & 2019.1.00618.S & 3-2 & $344 \times 268$ & \citet{alonso23} \\	
		NGC\,2992 &  2017.1.00236.S  & 2-1  & $184 \times 127$ & ALMA Archive \\
		NGC\,5728 &  2015.1.00086.S  & 2-1 & $562 \times 486$ & \citet{shimizu19} \\				
		NGC\,5135 &  2013.1.00243.S  & 2-1 & $308 \times 222$ & \citet{sabatini18} \\				
		\hline
		\hline
	\end{tabular}
	\label{alma_data}
\end{table*}

ALMA submillimetre spectroscopy of the low-excitation CO lines traces the distribution of circumnuclear cold gas in all our targets. 
We obtained moment0 maps of either the CO 2-1 or CO 3-2 lines, choosing observational datasets that offered comparable or better spatial resolution than the F1000W {\it JWST} images. 
Details of these datasets are given in Table \ref{alma_data} and references can be found in the discussion for individual targets in Section \ref{mwcomps}. For three targets without maps in literature sources, we used standard pipeline-processed outputs from the ALMA Science Archive to generate the maps.  

\subsection{{\it Spitzer}/IRS spectroscopy}

In the era before \jwst, the gold standard for spectroscopy over the full MIR wavelength range was set by the Infra-red Spectrograph (IRS) aboard the {\it Spitzer} Space Telescope. These spectra still serve as a major resource for large sample studies of AGN properties in the MIR, at least till \jwst\ can build a substantial archival legacy. Therefore, it is valuable to use our imaging data to understand the improvements that \jwst\ offers over IRS, particularly for constraints on the unresolved torus emission in our AGN (Sections \ref{nuclear_seds} and \ref{extlightfracs}).

We also use the IRS spectra to calculate a power-law approximation for the MIR SED shape of the AGN nuclear emission, as an input to the PSF modelling procedure discussed in Section \ref{psfs}.

Full details about the archival IRS spectroscopy used in this work can be found in Appendix \ref{irs_spectra}.

\section{Analysis methods}

\subsection{Emission line contamination} \label{emlinecont}

The wide-band MIRIM filters used in our \jwst\ programme span a number of bright emission lines associated with AGN NLRs and central starbursts. From work with the MIRI medium-resolution spectrometer (MRS), we know that line emission can influence the structures that we see in the images of our AGN targets, particularly in the bright parts of their NLRs \citep{haidar24, campbell25}. This `contamination' affects  measurements of dust continuum, the key science goal of the programme.

The following emission lines are key contaminants because of their intensity in AGN spectra: in F1000W, [S IV]$\lambda 10.51$ \mics; in F1500W, [Ne V]$\lambda 14.32$ \mics, [Ne III]$\lambda 15.56$ \mics; in F1800W and F2100W, [S III]$\lambda 18.71$ \mics. The F560W filter does not cover any strong lines.

In \citet{campbell25}, building upon a preliminary method developed in \citet{haidar24}, we presented a detailed exploration of emission line contamination estimation and mitigation strategies for our MIRIM images, based on techniques honed on MIRI/MRS spectroscopy and the use of multiple emission-line spatial templates. We show our best estimate of the degree of contamination in the F1000W images in the lower right subpanel of Figures \ref{ngc4388} -- \ref{ngc5135}. Most of our targets show regions with modest contamination at the level of 30-40\%, with small localised regions where it can exceed 50\%, particularly in NGC\,4388 and NGC\,5728. In Section \ref{cmap_results}, we present regional SEDs from the JWST photometry. For the three objects with existing MRS spectroscopy (NGC\,3081, NGC\,7172, NGC\,5728), where accurate contamination estimates can be made, we also show the contamination-corrected photometry. 

\subsection{MIRIM colour maps} \label{cmaps_prep}

Colour maps between different bands offer valuable insight into the spectral variation of the MIR emission across the central 28'' of our targets. 

While the five available MIRIM bands offer a number of permutations for pairwise colour maps, for the purposes of this work, we restrict our presentation to just two colours: F560W-F1000W and F1000W-F1500W. We avoid pairs of bands that are widely separated in wavelength because the large differences in PSF lead to artefacts in the colour maps. From examination, we found very little systematic colour variation in the F1500W-F1800W and F1800W-F2100W maps, so we choose not to present them here.

Using the model PSFs described in Section \ref{psfs}, we developed image convolution kernels to match the resolution of the short-wavelength image to that of the long-wavelength image in each colour filter-pair. The kernels were constructed using the PSF matching functionality of the PHOTUTILS package, based on the methods outlined in \citet{aniano11}. We used a split cosine-bell window function to avoid ringing effects on the kernels from high-frequency image components in the model PSF images, such as image edges. The parameters of the window function were varied to minimise the difference between a simulated PSF, created by convolving the short-wavelength PSF model with the kernel, and the known long-wavelength PSF model.

It is worth noting that none of the conclusions we derive from our examination of the colour maps in Section \ref{cmap_results} is dependent on the accuracy of the PSF-matching procedure. However, without matching the PSFs, visually-distracting ring-like artefacts form around compact features in colour maps.

When presenting the colour maps, we also mask out native columns and rows of the JWST images that passed within $0\farcs162$ of the nucleus. This simple excision, based on the half-width at half-maximum of the F1000W PSF, removes the worst residual effects of the poorly-modelled cruciform pattern from the nuclear point source.

\subsection{MIRIM photometry of the nuclear source} \label{sed_meas}


We examine the SED of the nuclear point source in each of our targets and place it in the context of dust emission constrained from NIR observations, ground-based and space-based MIR observations, and against the predictions from widely-used models of AGN tori.

We use two approaches to construct MIR SEDs of the nuclear point source. The first method is to perform photometry in our images over a small circular aperture $0\farcs4$ in diameter, a little larger than the FWHM of the F1000W PSF. We use aperture corrections calculated from the model PSFs (Section \ref{psfs}) to scale the aperture fluxes in each band to the total fluxes expected from the nuclear point source, under the assumption that all the flux in the aperture can be attributed to the point source. The statistical error on these estimates is miniscule, but they should be treated as formal upper limits to the true point source flux because any MIR emission from a resolved component is not removed.

Our second approach draws upon the point source modelling outlined in Section \ref{psf_subtraction}. This procedure yields a direct estimate of the flux of the nuclear point source in each band, separated from the emission from the underlying galaxy. The primary advantage of this approach is that it relaxes the assumption that all the nuclear light comes from a point source, which may not be accurate in the objects with bright extended MIR emission components. On the other hand, the modelling relies on the validity of a S\'ersic function for the azimuthally-averaged light profile of the extended emission, which could be inaccurate, especially when extrapolated towards small nuclear radii. We estimate a very conservative error on the modelled point source fluxes by taking the most negative residual value within the EER80 and assuming that it defines the maximum possible deviation of the normalisation of the fitted point source. This error is shown in Figures \ref{all_sedfig1} \& \ref{all_sedfig2}. 

We present our estimates of the nuclear fluxes using both these approaches in Table \ref{nuclear_fluxes}. We discuss the resultant SEDs and compare them to expectaations in Section \ref{nuclear_seds}.

We also include estimates of the NIR dust emission in our targets, derived in two ways. The first comes from HST images at 1.6 \mics\ or 2 \mics, many of which show a nuclear point source. We measure the fluxes from these images in a circular aperture of $0\farcs26$ diameter centred on the nucleus, and apply a fixed aperture correction of $2.02$ to these fluxes, consistent with the known PSF of HST in the F160W band. This yields an estimate of the maximal NIR dust nuclear emission of our targets, valid only if there is no substantial stellar contamination, and no extended hot dust.

A second estimate of the NIR dust emission comes from the compilation of nuclear continuum fluxes from \citet{burtscher15}, which are calculated from the dilution of stellar features in SINFONI spectroscopy at 2.2 \mics. While these fluxes are free of the effects of stellar emission in the nucleus, they are subject to the larger uncertainties in spectrophotometric calibration that can affect SINFONI data.

\section{Results} \label{results}

\subsection{Nuclear SEDs} \label{nuclear_seds}

\begin{table*}
	\caption{Nuclear point-source and extended flux density measurements. Extended fluxes are measured from PSF-subtracted images (Section \ref{psf_subtraction})}	\begin{tabular}{llllll}
		\hline
		\hline
		Type of estimate & F560W (mJy) & F1000W (mJy) & F1500W (mJy)  & F1800W (mJy) & F2100W (mJy)  \\
		\hline
		\hline
		\multicolumn{6}{l}{NGC\,4388} \\
		\hline 
		Corrected aperture: & $ 61.75$ & $ 79.06$ & $348.44$ & $523.75$ & $729.95$ \\
		PSF model: &  $ 40.96\pm10.37$ & $ 74.03\pm11.92$ & $319.28\pm 7.11$ & $477.89\pm10.25$ & $667.05\pm16.07$ \\
		5 arcsec: &  88.88 & 123.46 & 573.13 & 831.91 & 1079.21  \\
		IRS SL/LL slit: &  82.84 & 113.79 & 646.67 & 948.55 & 1230.57  \\ 
		\hline 
		\multicolumn{6}{l}{NGC\,3227} \\
		\hline 
		Corrected aperture: & $ 87.60$ & $246.27$ & $558.53$ & $887.70$ & $1003.72$ \\
		PSF model: & $ 60.75\pm16.16$ & $238.10\pm68.28$ & $496.88\pm13.60$ & $849.25\pm 6.62$ & $1028.22\pm182.63$ \\
		5 arcsec: & 140.37 & 306.83 & 758.29 & 1114.56 & 1211.01  \\ 
		IRS SL/LL slit: & 131.31 & 292.11 & 821.99 & 1218.53 & 1323.58  \\ 
		\hline 
		\multicolumn{6}{l}{ESO 428-G14} \\
		\hline 
		Corrected aperture: & $ 35.37$ & $ 99.98$ & $368.29$ & $496.66$ & $550.29$ \\
		PSF model: & $ 24.94\pm 6.89$ & $ 70.88\pm 2.91$ & $251.98\pm10.66$ & $293.31\pm14.07$ & $226.78\pm13.34$ \\
		5 arcsec: &  59.49 & 177.15 & 705.42 & 983.10 & 1088.87  \\ 
		IRS SL/LL slit: &  53.88 & 164.32 & 772.09 & 1094.09 & 1217.81  \\ 
		\hline 
		\multicolumn{6}{l}{NGC\,3081} \\
		\hline 
		Corrected aperture: & $ 51.44$ & $145.30$ & $294.85$ & $395.52$ & $455.21$ \\
		PSF model: & $ 43.06\pm 2.73$ & $127.94\pm 2.46$ & $286.85\pm15.75$ & $347.74\pm 6.72$ & $342.31\pm12.20$ \\
		5 arcsec: &  66.15 & 190.95 & 512.49 & 699.08 & 773.02  \\ 
		IRS SL/LL slit: &  62.55 & 184.02 & 551.45 & 760.34 & 842.06  \\ 
		\hline 
		\multicolumn{6}{l}{NGC\,7172} \\
		\hline 
		Corrected aperture: & $192.09$ & $ 51.85$ & $314.43$ & $232.80$ & $272.82$ \\
		PSF model: & $142.62\pm41.70$ & $ 50.83\pm14.59$ & $293.30\pm 4.12$ & $198.92\pm 1.51$ & $245.99\pm 2.97$ \\
		5 arcsec: & 247.22 &  68.07 & 391.24 & 299.09 & 336.26  \\ 
		IRS SL/LL slit: & 234.68 &  62.35 & 447.48 & 360.14 & 397.66  \\ 
		\hline
		\multicolumn{6}{l}{NGC\,2992} \\
		\hline 
		Corrected aperture: & $ 54.44$ & $147.30$ & $360.37$ & $496.27$ & $560.04$ \\
		PSF model: & $ 32.93\pm 8.67$ & $138.44\pm23.90$ & $332.47\pm 5.93$ & $466.30\pm 3.61$ & $523.98\pm 8.26$ \\
		5 arcsec: &  83.97 & 184.20 & 465.51 & 608.13 & 652.59  \\ 
		IRS SL/LL slit: &  76.52 & 174.27 & 543.52 & 723.84 & 784.40  \\
		\hline
		\multicolumn{6}{l}{NGC5728} \\
		\hline 
		Corrected aperture: &$ 10.83$ & $ 12.78$ & $112.10$ & $144.89$ & $202.33$ \\
		PSF model: & $  7.69\pm 0.35$ & $  6.86\pm 0.55$ & $ 79.49\pm 4.06$ & $101.76\pm 4.56$ & $174.19\pm 5.35$ \\
		5 arcsec: &  23.43 &  32.97 & 177.17 & 226.95 & 287.80  \\ 
		IRS SL/LL slit: &  19.64 &  26.98 & 263.51 & 356.74 & 417.73  \\ 
		\hline 
		\multicolumn{6}{l}{NGC\,5135} \\
		\hline 
		Corrected aperture: & $ 26.28$ & $ 47.42$ & $157.14$ & $223.02$ & $307.56$ \\
		PSF model: & $ 19.86\pm 6.25$ & $ 46.32\pm13.52$ & $160.38\pm44.86$ & $213.94\pm11.18$ & $265.81\pm26.77$ \\
		5 arcsec: &  71.20 & 142.91 & 436.43 & 694.99 & 943.53  \\ 
		IRS SL/LL slit: &  52.83 & 105.47 & 587.18 & 948.66 & 1268.58  \\ 
		\hline  
		\hline
	\end{tabular}
	\label{nuclear_fluxes}
\end{table*}

\begin{figure*}
	\centering
	\includegraphics[width=0.95\textwidth]{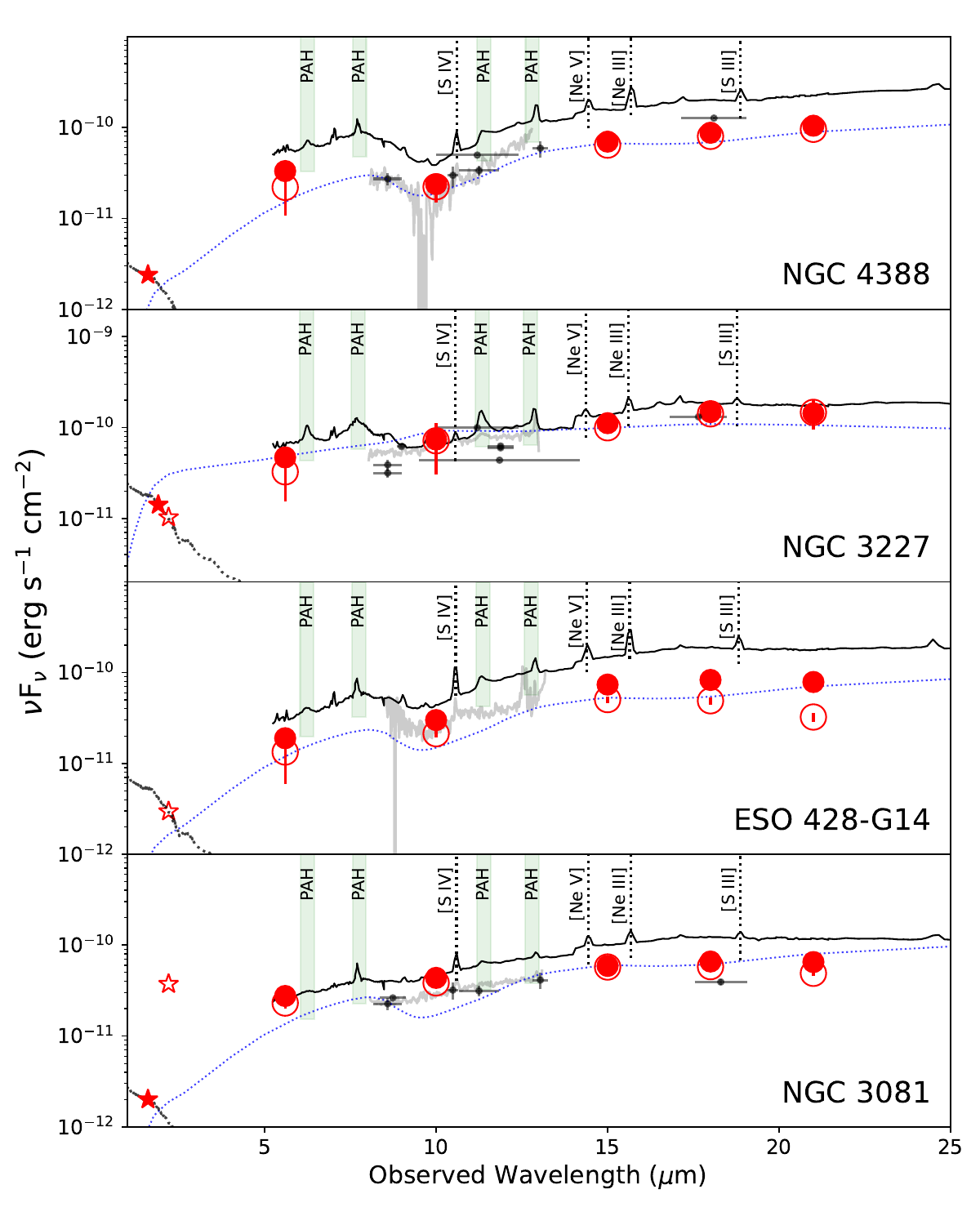}		
	\caption{Nuclear SEDs of AGN from our sample. Red filled points show SEDs from aperture-corrected fluxes, while open circles show SEDs of the central point source modelled as a PSF. The error bars on the open red points are a conservative estimate of the systematic uncertainty from the point-source subtraction method. The dotted blue line shows a characteristic clumpy torus spectral model from \citet{garcia-bernete19}. Small black points show various ground-based flux measurements, while the grey line shows a selected ground-based spectrum. The solid black line is the {\it Spitzer}/IRS low-resolution spectrum.  Spectral features that have been labelled can potentially contaminate the continuum in the MIRI filters bandpasses. Red star points show NIR photometric estimates of the nuclear flux: open stars from \citet{burtscher15} and filled stars from HST images. The dotted black line is a stellar population template scaled to the HST NIR flux, showing the maximal amount of stellar light that may contaminate the nuclear SEDs. }
	\label{all_sedfig1}
\end{figure*}

\begin{figure*}
	\centering
	\includegraphics[width=0.95\textwidth]{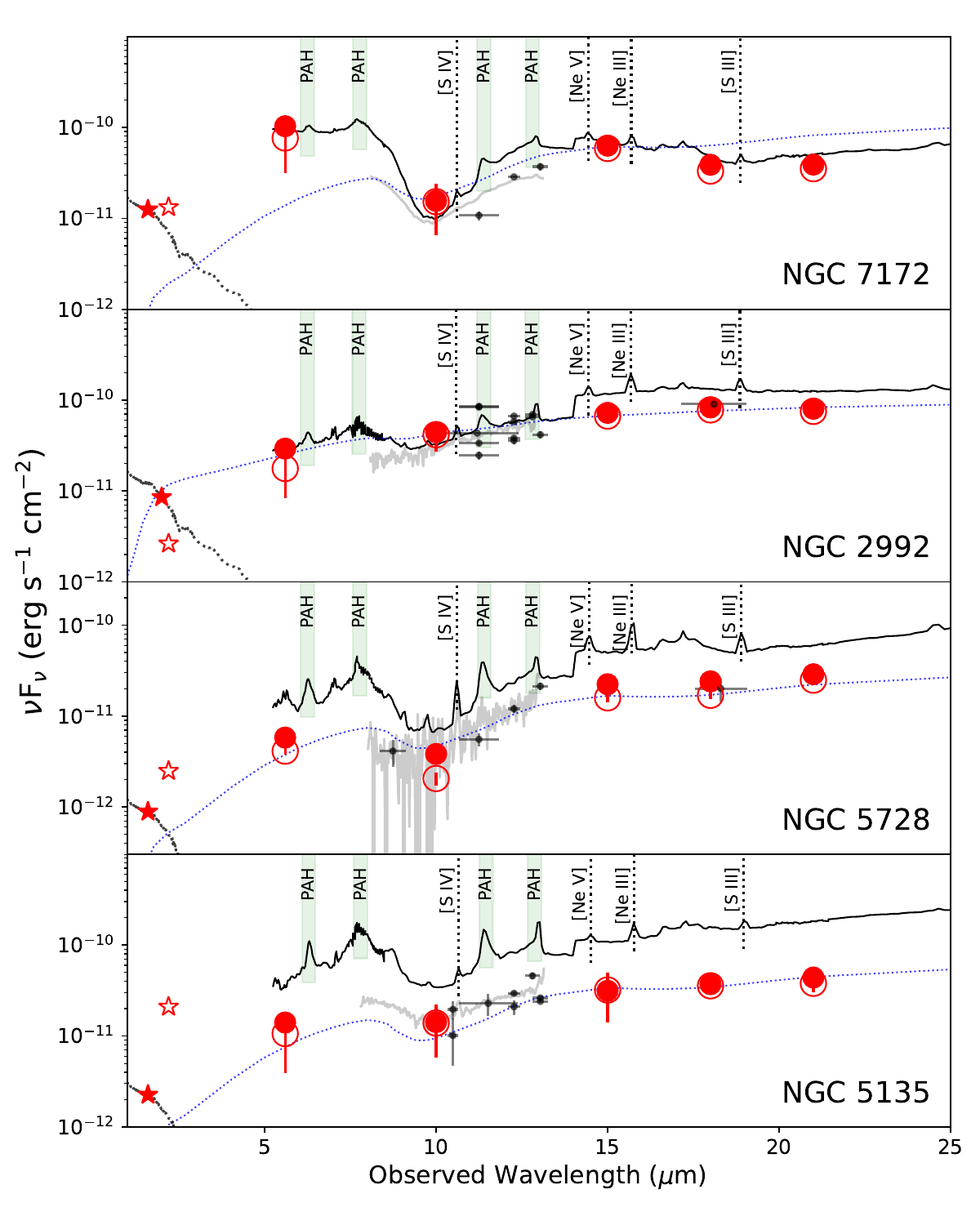}	
	\caption{Same as for Figure \ref{all_sedfig1}}
	\label{all_sedfig2}
\end{figure*}

In Figure \ref{all_sedfig1} \&  \ref{all_sedfig2}, the nuclear NIR-to-MIR SEDs for all 8 targets are shown (see Section \ref{sed_meas} for details). In each panel, we also include a reference SED derived from a work of \citet{garcia-bernete19}, in which the nuclear SEDs of Seyfert galaxies were fit with the clumpy torus models of \citet{nenkova08}. 
Most of our targets are classified as Seyfert 2s, and we compare our nuclear photometry of these galaxies to the median Seyfert 2 model SED from \citet{garcia-bernete19}. NGC\,3227 and NGC\,2992 are classified as Seyfert 1.5 and 2i respectively: for these AGN, we plot the median Seyfert 1 and intermediate Seyfert model SEDs from \citet{garcia-bernete19}. In all cases, we have normalised the model SED to the PSF-scaled flux in the F1500W band.

We plot a model spectrum of a typical bulge stellar population scaled to the HST NIR photometric point, as a guide to the maximum emission we may expect from stars in the NIR-to-MIR. Because of the strongly dropping shape of the stellar photospheric SEDs with increasing wavelength, stellar emission in even the shortest wavelength JWST band
(F560W) is at least an order of magnitude lower that what we measure.
It is clear that all the unresolved emission in the JWST bands can be safely attributed to the dust continuum from the nucleus\footnote{Emission line contamination in the nuclear light is negligible \citep{campbell25}}. 

We find that, particularly at wavelengths $\geq 15$ \mics, the median clumpy torus models are a good match to the shape of the empirical nuclear SEDs. At shorter wavelengths, we find differences between the model curves and the data points, but these can easily be understood as the effects of a different torus obscuration or dust covering factor compared to a median Seyfert nucleus.

Only the SED of NGC\,7172 deviates strongly from the equivalent model SED, primarily at short wavelengths, where there is a substantial hot dust excess which boosts the flux in the NIR and F560W band. The atypical SED shape is also seen in the archival IRS spectrum; with the JWST photometry, we can confirm that it is not caused by off-nuclear emission entering the IRS slit. Indeed, NGC\,7172 has the second highest nuclear source contrast in the sample, after NGC\,3227, and almost all the flux in the IRS spectrum can be attributed to the JWST point source. The very strong silicate absorption in NGC\,7172 has been attributed to a dense dust screen in the edge-on host disk \citep[e.g.,][]{garcia-bernete24a}. 

In each panel, we show the corresponding low-resolution optimally-extracted {\it Spitzer}/IRS spectrum (Section \ref{irs_data}). As expected, the IRS spectra are always brighter than, or at least as bright as, the JWST nuclear photometry at the same wavelengths, indicating the presence of considerable non-nuclear emission within the much larger apertures used to extract these spectra. We examine the amount of extended emission captured by IRS spectroscopy in Section \ref{extlightfracs}.

The figure for each object also includes a set of ground-based high-resolution MIR photometry compiled from the SASMIRALA database\footnote{\url{dc.g-vo.org/sasmirala/q/cone/info}} \citep{asmus14}, and a representative ground-based MIR spectrum of the nucleus. The higher spatial resolution of the ground-based data ($\approx 0\farcs1$) compared to {\it Spitzer} allows better isolation of the nuclear emission. Consequently, the ground-based fluxes are also fainter than the IRS spectrum in most cases; when they are brighter, there is disagreement between multiple flux estimates, suggesting issues with the ground-based data.

For all galaxies, there is good general agreement between the ground-based and JWST/MIRI estimates of the nuclear fluxes. This is an important validation of the approaches we have taken to estimate the nuclear fluxes from these images. We can proceed with the knowledge that both the normalisation and the shape of the nuclear SEDs are good representations of the central emission from the AGN, at least over the wavelengths spanned by JWST/MIRI.

Future work from our team will focus on fitting the nuclear emission of the sample, spanning from optical through MIR wavelengths, using a full suite of various physically-motivated models, including those that include unresolved dusty wind components \citep[e.g.][]{hoenig17}. We reserve further treatment of the nature of the nuclear emission to these upcoming studies.

\subsection{Extended light fractions in Spitzer/IRS spectra }  \label{extlightfracs}

From Figures \ref{all_sedfig1} \& \ref{all_sedfig2}, it is clear that the point source fluxes measured from the JWST images can be many factors fainter than the fluxes inferred from central IRS spectra of our Seyfert targets. This implies that, for local Seyfert galaxies, the IRS slit width apertures ($3\farcs7$ for the Short-Low (SL) modules and $10\farcs7$ for the Long-Low (LL) modules) encompass a substantial amount of extended dust and line emission which cannot be attributed to the torus. Here, we assess the amount of extended emission using our JWST images and point-source flux estimates, to serve as a guideline for studies that rely on {\it Spitzer}/IRS spectroscopy to study AGN nuclear regions.

We compute the total fluxes from the JWST images in two representative apertures: a fixed circular aperture of $5\farcs0$ diameter\footnote{The approximate diffraction-limited PSF width of {\it Spitzer} at 15 \mics} for all 5 bands, and a square aperture equal to the IRS SL/LL slit for bands with central wavelengths less than/greater than 14 \mics, the transition wavelength between the SL and LL spectral modules. We use the point source modelling results from Section \ref{psf_subtraction} to subtract the nuclear flux from these aperture fluxes, giving us the total amount of extended flux in each band and aperture. The fluxes of extended light in the large apertures are listed in Table \ref{nuclear_fluxes}.

Between 20\% and 70\%  of the IRS flux at 10 \mics\ can arise from extended emission. As may be expected, there is a correlation betwen the fraction of extended light and the prominence of extended emission in the \jwst\ images of 5'' scales. As we discuss later, in most of our targets, the extended emission within the inner several arcseconds arises from dust that is in the NLR, directly under the influence of the AGN's radiation field. The extended light fluxes we have calculated are a rough measure of the prominence of the NLR emission relative to the torus emission in the sample.

We can compare these fractions to the relative flux arising from NLR emission reported in the literature from IRS-based studies. \citet{mor09} used spectral decomposition techniques to separate NLR, torus and hot dust emission in a sample of PGQSOs, finding typical NLR light fractions of $\approx 15$\% at 10 \mics\ and $\approx 40$\% at 18 \mics, integrated over scales of several kpc. While we do find similar fractions among our Seyferts at longer wavelengths, at 10 \mics\ we find a substantially higher fraction of extended light than reported in \citet{mor09}. At face value, this would imply a much redder NLR dust emission SED across the MIR in PGQSOs ($L_{bol} \approx 10^{45\textrm{-}46}$ \ergs) compared to Seyferts ($L_{bol} \approx 10^{43\textrm{-}44}$ \ergs). However, it is likely that the relative contribution from star formation-heated dust is also higher among Seyferts because of their lower AGN luminosities, necessitating a decomposition of NLR emission from other sources in the circumnuclear region. Future studies of the SEDs of extended dust emission will explore the intrinsic properties of NLR dust in more detail.

\subsection{MIRIM colour maps: general trends} \label{mir_colours}

\begin{figure*}
	\centering
	\includegraphics[width=\textwidth]{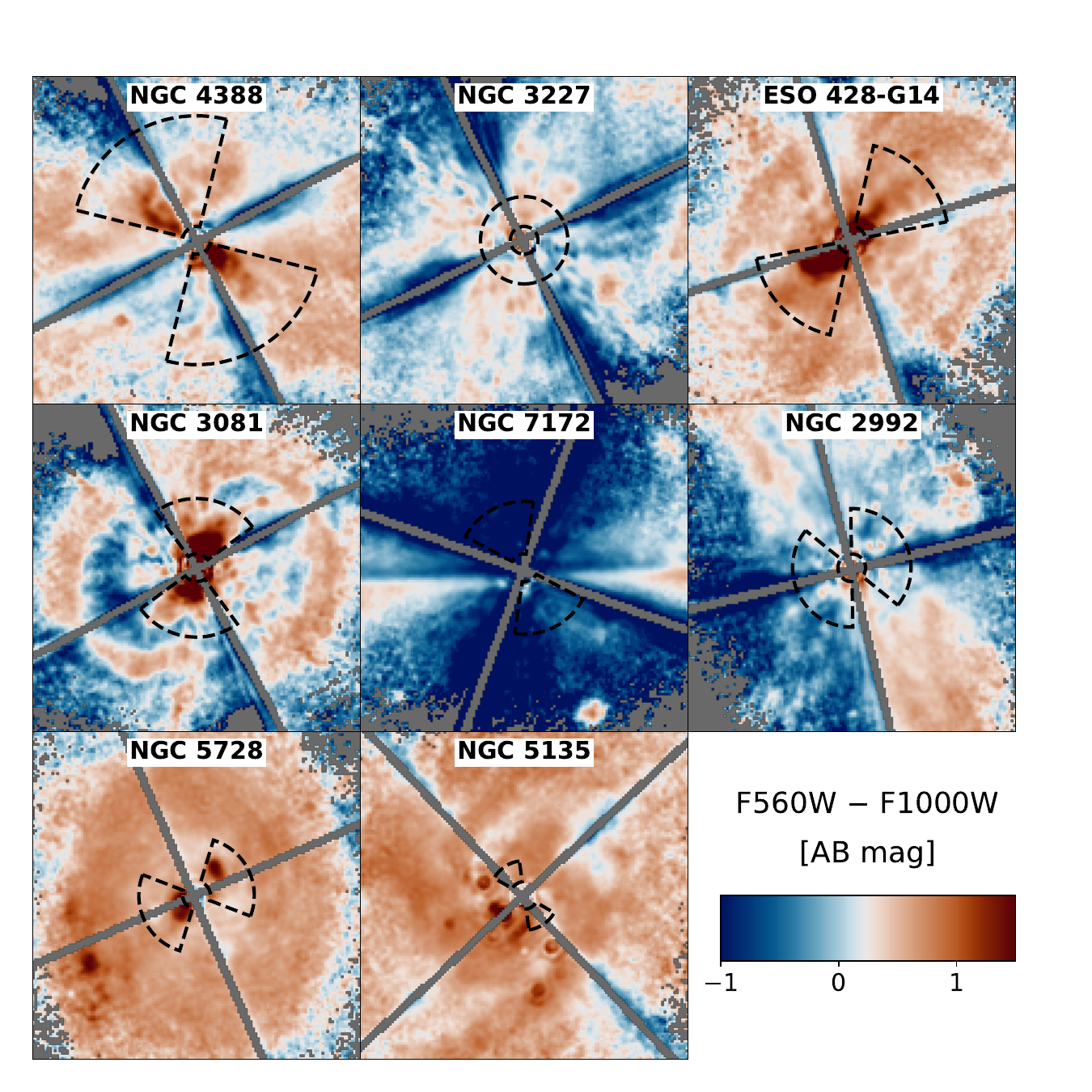}	
	\caption{Maps of F560W-F1000W colour for all targets, constructed from PSF-subtracted images that are then PSF-matched. Each panel covers 15'' on a side centered on the nucleus. The galaxies are arranged in order of distance from the top left. The dashed wedges delineate the approximate shape of the known ionisation cones in these AGN. NGC\,3227 has an ionisation cone aligned close to the line of sight, therefore a circle is used to show the approximate extent of the brighter NLR. The outer extent of the wedges mark a projected distance of 500 pc from the nucleus. The inner dashed arcs or circles give the EER80 of the F1000W PSF, to delineate regions within which PSF subtraction residuals can be important. The grey cross-shaped mask excludes pixels which are influenced by the PSF cruciform pattern, particularly strong in the F560W and F1000W images (see Section \ref{psf_subtraction}).
	}
	\label{f560w-f1000w}
\end{figure*}

\label{cmap_results}

We now turn our focus to the structure and photometric properties of the extended emission in our targets, as revealed by the colour maps (Section \ref{cmaps_prep}).

In Figure \ref{f560w-f1000w}, we show maps of the F560W-F1000W colour for all 8 galaxies in our programme. The scaling that we have used is the same for all the maps so that the galaxies can be easily intercompared. In each of the maps, we have also shown the approximate shape of the NLR ionisation cone as determined from the structure of the [Si VI]$\lambda 1.96$ \mics\ emission. The cones are drawn to a fixed extent of 500 pc from the nucleus. The inner radius of the cones indicate the EER80 of the PSF of the F1000W filter ($0\farcs64$). Any structure within this radius may be significantly affected by PSF subtraction errors or kernel mismatches (see Sections \ref{psf_subtraction} and \ref{cmaps_prep} for details).

The most prominent unmasked features in the colour maps are the central red structures, seen in most of the targets. These structures are aligned with the ionisation cones, and delineate the inner NLRs, with radial sizes of 200--300 pc. They are particularly prominent in NGC\,4388, ESO 428-G14, NGC\,3081, and NGC\,5728, all of which have large, bright NLRs. The structures are not evident in NGC\,3227 and NGC\,2992. In the case of NGC\,5135, several bright knots are associated with features in the circumnuclear starburst, but we find no clear correpondence with the NLR.  

We also clearly detect the circumnuclear dusty disk in all our targets, particularly in the F560W images. In  ESO 428-G14, NGC\,3081, NGC\,5728 \& NGC\,5135, the disk shows pronounced spiral structure \citep[e.g.][]{haidar24}. NGC\,4388, NGC\,7172, and NGC\,2992 have prominent, edge-on disks which show up clearly in all the colour maps. In contrast to the red F560W-F1000W colours of the NLRs or starburst regions, the extended star-forming disks are bluer, with a relatively flat SED between 5 and 10 \mics\ (F560W-F1000W $\approx 0.5$).

\begin{figure*}
	\centering
	\includegraphics[width=\textwidth]{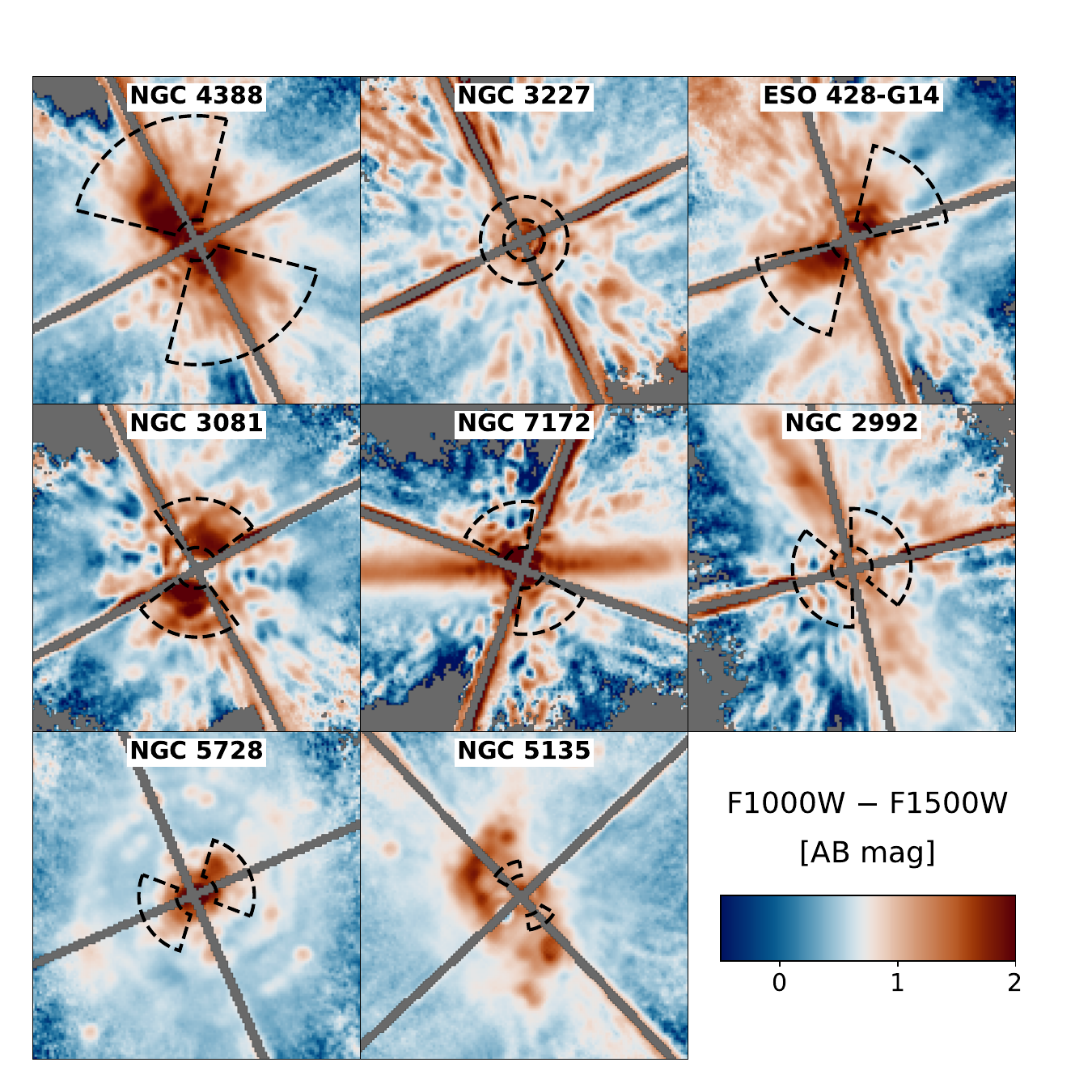}	
	\caption{Maps of F1000W-F1500W colour for all targets, constructed from PSF-subtracted  images that are then PSF-matched. Each panel covers 15'' on a side centered on the nucleus. The galaxies are arranged in order of distance from the top left. The dashed wedges delineate the approximate shape of the known ionisation cones in these AGN. NGC\,3227 has an ionisation cone aligned close to the line of sight, therefore a circle is used to show the approximate extent of the brighter NLR. The outer extent of the wedges mark a projected distance of 500 pc from the nucleus. The inner dashed arcs or circles give the EER80 of the F1500W PSF, to delineate regions within which PSF subtraction residuals can be important. The grey cross-shaped mask excludes pixels which are influenced by the PSF cruciform pattern, particularly strong in the F560W and F1000W images (see Section \ref{psf_subtraction}).
	}
	\label{f1000w-f1500w}
\end{figure*}

Colours maps based on F1000W-F1500W are shown in Figure \ref{f1000w-f1500w}. 
The disk and extended spiral arms are less pronounced in these maps. Instead, compact knots of star-formation in the disks stand out more clearly. Clear edge-on disks are seen in the NGC\,7172 and NGC\,2992 maps; interestingly, their apparent vertical thicknesses are narrower in these maps compared to F560W-F1000W.

In contrast to the extended galaxy dust, the regions within the AGN ionisation cones (or the starburst in NGC\,5135) are significantly redder in F1000W-F1500W, showing up as very prominent features within the inscribed ionisation cones. NGC\,3227 and NGC\,2992 are exceptions, but the NLRs in both AGN are compact and can be confused by diffraction residuals that remain in the maps.

\begin{figure*}
	\centering
	\includegraphics[width=\textwidth]{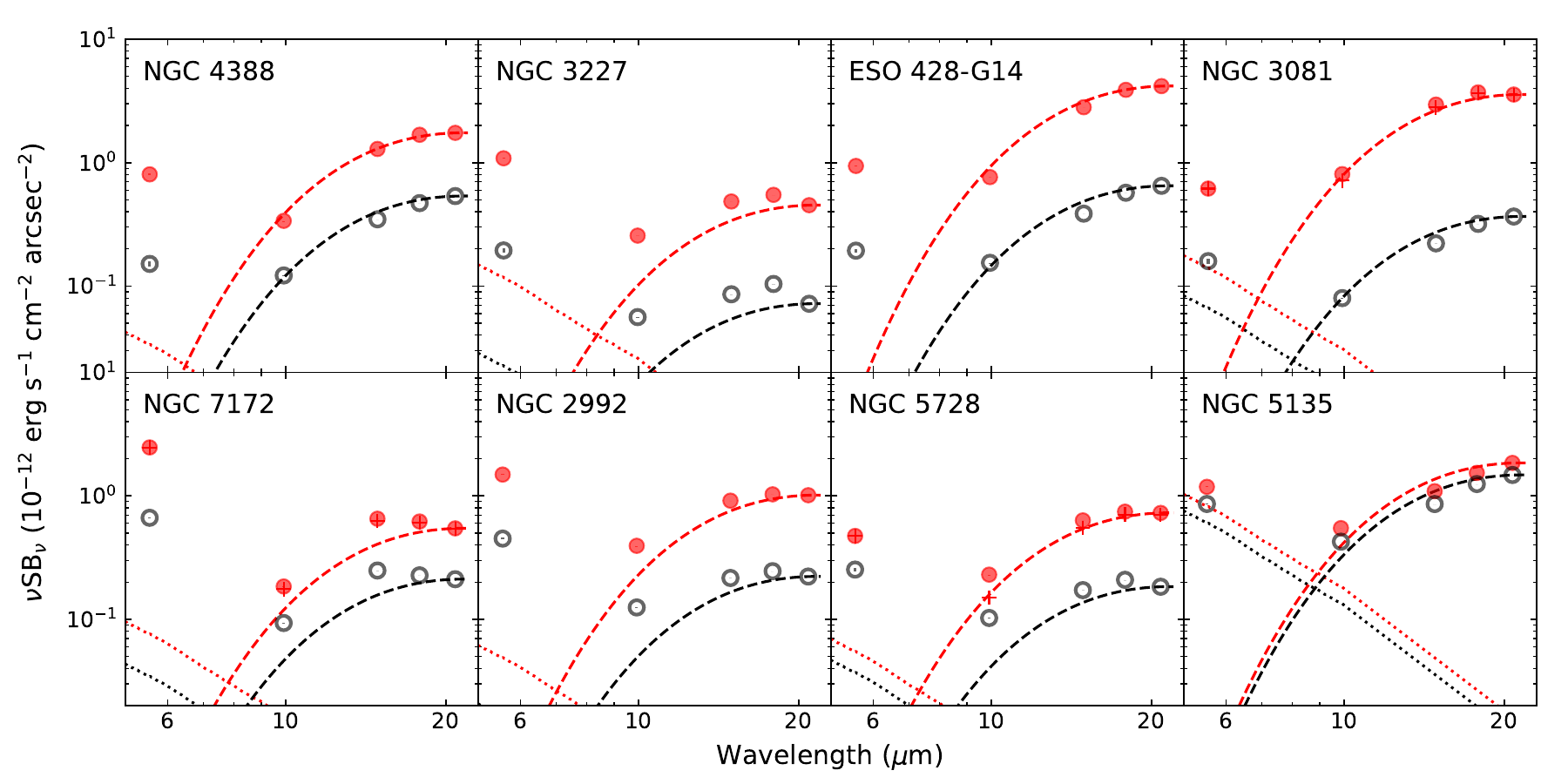}	
	\caption{SEDs of the extended emission within 500 pc of the nucleus (1 kpc for NGC\,5135) within the NLR (red filled points) and outside the NLR (black open points). For objects with MIRI spectroscopy, emission line-corrected fluxes are plotted as red crosses. The red and black dashed lines show the spectral shape of a blackbody with a temperature of 170 K, scaled to the F2100W flux of the corresponding SED. The red and black dotted lines show the maximal stellar contribution extrapolated from HST NIR photmetry. Statistical errors on JWST flux measurements are included, but are generally negligible. Note that the surface brightness of the emission within the NLR is consistently higher than the emission outside the NLR.
	}
	 \label{sed_500pc}
\end{figure*}

The strong red F1000W-F1500W colours of the AGN-influenced material in the NLR ionisation cones indicates a different MIR SED compared to material
that lies outside the cones, and hence, not illuminated by the AGN. We examine this in Figure \ref{sed_500pc}, where we plot the averaged SEDs of regions within 500 pc of the nucleus of the galaxies (1 kpc for NGC\,5135), comparing those within the NLRs with those outside.

To select pixels that lie within the NLRs, we used the [Si VI] flux maps of all our targets as a spatial mask. These maps were first reprojected to the image scale and orientation of the JWST images of the respective galaxies, and then a surface brightness threshold was manually determined to ensure that all parts of the NLR with bright line emission were encapsulated. These thresholds are indicated using a single contour level in the [Si VI] maps of Figures \ref{ngc4388} -- \ref{ngc5135}.

The emission line-based masks were used to identify pixels outside the NLR, and the inverse of the mask, corresponding only to regions with bright line emission, were used to identify pixels within the NLR. 
In addtion to the line masks, we also applied the cruiform pattern masks discussed above. 

From Figure \ref{sed_500pc}, we find that the surface brightness of emission within the NLR is always higher than that outside the NLR, by factors of $2$--$10$. The only exception is NGC\,5135, where both points within and outside the nominal NLR cone have very simular SEDs and surface brightnesses. This is because the MIR emission in the inner few kpc of this object is dominated by the starburst, with no evidence for excess emission along the axis of the NLR.   

The MIR SEDs of NLR regions are also generally steeper than those outside the NLR, rising more sharply towards longer wavelengths, particularly between 10 and 15 \mics. As a guide to the eye, we show in Figure \ref{sed_500pc} the SEDs of model blackbodies at a fixed temperature of 175 K, scaled to the F2100W flux of both NLR and non-NLR SEDs of each galaxy. While a single blackbody spectrum cannot reproduce any of the SEDs over the full 5--21 \mics\ range, one sees that the deviations from the reference blackbody model curves are larger for the points corresponding to the regions outside the NLR (black open circle points).

\subsection{Multiwavelength comparisons} \label{mwcomps}

In this Section, we examine the multiwavelength datasets of our targets in concert. 
We present our results in an object-by-object fashion, starting with a brief description of the known nuclear and circumnuclear properties of each target. We place the information gained from the colour maps, line contamination maps, and the multiwavelength comparisons into this context, and highlight a few key heuristic results for each galaxy.

\subsubsection{NGC\,4388}

\begin{figure*}
	\centering
	\includegraphics[width=\textwidth]{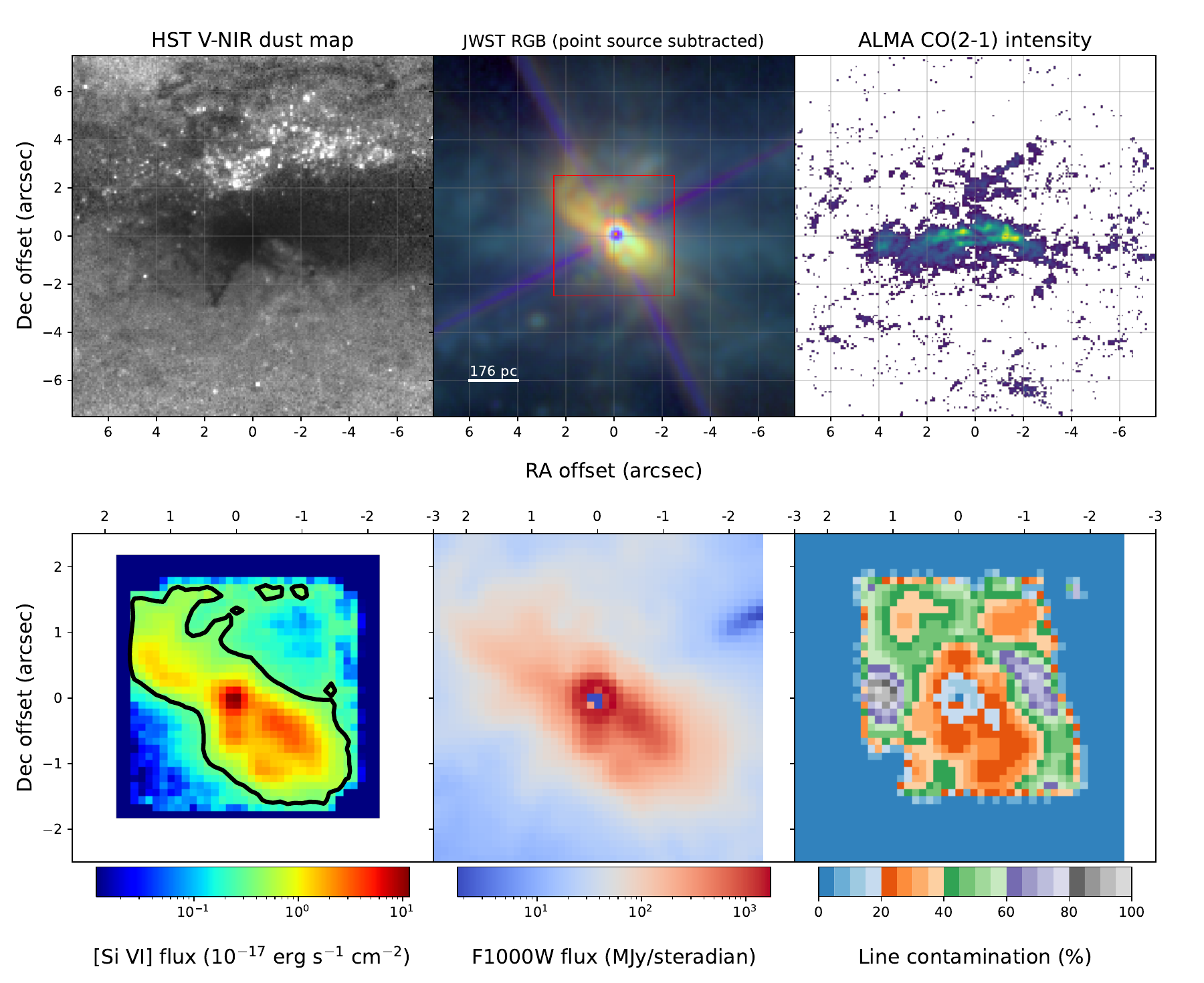}	
	\caption{Context figures for NGC\,4388 Top: Multiwavelength images covering the inner 15 arcseconds of the galaxy. The 2 arcsecond scale bar in the middle image is labelled by the projected physical scale at the distance of the system. All images are oriented equatorially with North to the top and East to the left. Darker regions of the {\it HST} dust map are more dust-obscured. Yellower colours of the CO map correspond to brighter line fluxes. The central RGB image follows the colour scheme of Figure \ref{all_RGB}.
	Bottom: Maps over the central 5'' of the galaxy, of the [Si VI]$\lambda$ 1.96 \mics\ flux (left), PSF-subtracted MIRIM F1000W flux, and line contamination fraction in the F1000W filter (right). The size of the region in these panels is shown by the red square in the middle panel of the top row. Note that the [Si VI] map may be smaller than the 5'' field. The thick black contour on the [Si VI ] map indicates the surface brightness threshold used to create an NLR spatial mask (see Section \ref{cmap_results}). 
	}
	\label{ngc4388}
\end{figure*}

As a gas-rich member of the Virgo cluster, NGC\,4388 (Figure \ref{ngc4388}) has been studied both for its X-ray bright AGN \citep[e.g.,][]{hanson90, elvis04} and for the interaction of its galaxy disk with the intra-cluster medium \citep{veilleux99a}. The host galaxy, an SB spiral with uncertain classification \citep{rc3}, is almost edge-on, and a dense dust lane obscures the nucleus at optical wavelengths \citep[E(B-V)$\approx 1.9$;][]{rodriguezardila17}. The kinematics of the ionised gas indicates bar streaming motions, indicative of a bar potential underlying the boxy bulge of the galaxy \citep{veilleux99b}.

Though optical imaging spectroscopy shows a one-sided ionisation cone \citep{stoklasova09}, the bipolar nature of the NLR is revealed with NIR integral-field (IF) spectroscopy \citep{greene14}. There is close morphological correspondence between the structure of the NLR and that of the radio jet as seen in high-resolution VLA images \citep{hummel91,sargent24}. 

From the top panels in Figure \ref{ngc4388}, we can contrast the optical, MIR, and sub-mm view of the cool phases in NGC\,4388. The prominent edge-on disk is seen in the optical dust map (left top panel) in the form of dark dust lanes, lined with a smattering of bright star-forming regions. The CO map (right top panel) reveals the more centrally concentrated denser molecular gas that surrounds the nucleus, which is invisible in the optical due to the overlying dust further out in the disk. This edge-on molecular disk is about 700 pc in radius. Though less pronounced in the middle top panel, the disk is also visible in the MIR image. Much of its substructure can be seen in the dust emission, and the disk itself can be traced further out to beyond the inner 1.5 kpc.

The most prominent structure in the JWST images is an elongated set of features that lie along a PA of $\approx 45^{\circ}$. This structure is brought out clearly in the F1000W-F1500W map for the galaxy in Figure  \ref{f1000w-f1500w}, where it shows uniformly red colours. This is a prime example of the tendency of AGN-influenced MIR emission in all of our targets to show such a steep rise in flux from 10 \mics\ to 15 \mics.

The bright extended structure matches up in detail with the high-excitation ionised gas structures seen in circumnuclear optical and NIR imaging spectroscopy of NGC\,4388 by earlier works \citep{falcke98, stoklasova09, greene14}. Optical emission lines such as [O III]$\lambda 5007$ show a strongly one-sided structure extending 700 pc towards the south-west, with sharp edges reminiscent of a cone. In the NIR, the north-eastern extension of the emission line region is revealed in lines such as Br$\gamma$ and [Si VI]$\lambda 1.963$ \mics\ \citep{greene14}, indicating that the northern part of the NLR is both obscured by the intervening dust lane and aligned away from our line of sight. An outflow is visible in the emission line gas, redshifted in the northeast and blueshifted in the southwest with respect to systemic, consistent with the geometry of the NLR laid out above.

In \citet{haidar26}, we have demonstrated that the bright MIR emission aligns very well with the structure of the high-ionisation emission lines seen in NGC\,4388 . Particularly interesting is the feature seen extending to the north of the nucleus in the CO map, which bends towards the northwest after about 2 arcseconds. The feature has a counterpart in the MIRI images. While most of the CO-emitting disk has blue F1000W-F1500W colours, this tendril of CO matches a feature in the MIR with redder colours consistent with the rest of the AGN-influenced NLR.

\subsubsection{NGC\,3227}

\begin{figure*}
	\centering
	\includegraphics[width=\textwidth]{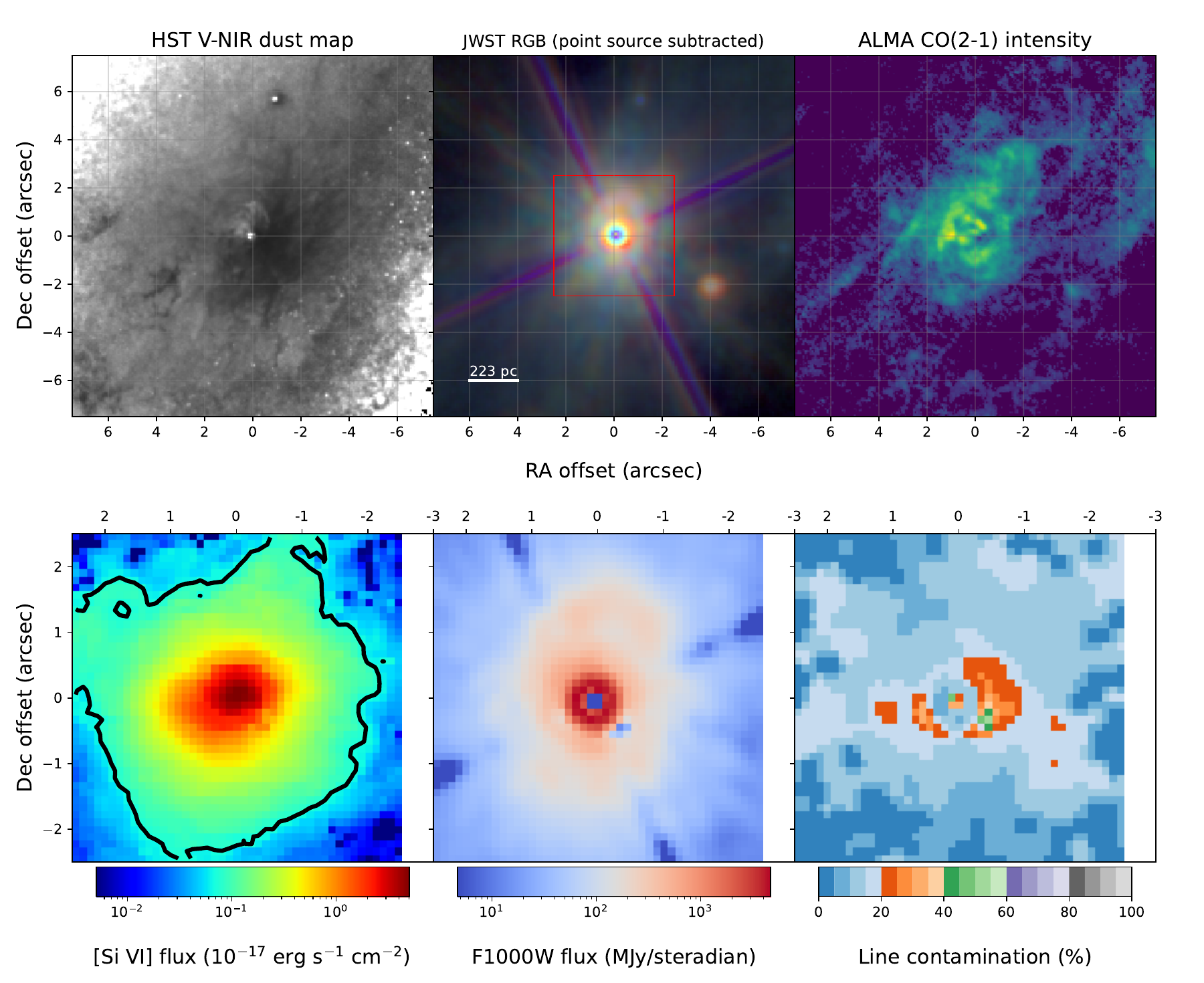}
	\caption{Context figures for NGC\,3227. See the caption for Figure \ref{ngc4388} for explanations for each of the panels.
	}
	\label{ngc3227}
\end{figure*}

Part of a strongly interacting pair of galaxies, NGC\,3227 (Figure \ref{ngc3227}) is traditionally classified as an SAB spiral, with a disk inclined by $\approx 52^{\circ}$. More recent NIR imaging studies reveal a kpc-scale bar \citep{mulchaey97} which is responsible for streaming motions that carry gas down to a circumnuclear ring with a radius of $\approx 5"$ \citep{barbosa06, davies14}. ALMA studies of the inner 1'' show a nuclear disk in submillimetre continuum and molecular gas, which may be inclined with an axis further from the line of sight than the larger scale galaxy disk \citep[i.e., a ``warped'' nuclear disk,][]{alonso-herrero19}. In optical emission lines, the NLR shows a strong asymmetry and an extension towards the north-east (PA$\sim 30^{\circ}$). However, in NIR emission lines, the NLR, as traced by the [Si VI] line, is quite round and symmetric (Figure \ref{ngc3227}). The difference is likely due to strong dust extinction to the south-west of the nucleus, tracing the nearer edge of the bar-mediated dust lane, which modifies the apparent distribution of the optical emission lines and hides the south-western extension of the NLR. Nevertheless, the orientation of the nuclear disk and the direction of outflowing gas seen in ionised and molecular gas \citep{davies14, alonso-herrero19} indicates a primary axis for the AGN's illumination in the PA range of $30^{\circ}$--$40^{\circ}$.

A one-sided radio jet extending $0\farcs4$ along a PA$\sim-10^{\circ}$ has been imaged with MERLIN. The different PA of the jet and the NLR has been taken as evidence for a physical misalignment between the axis of the SMBH accretion and the radio ejection \citep{mundell95}.   

NGC\,3227 has the strongest MIR nuclear point source contrast of our sample, and the PSF features that remain, even after optimised subtraction, indicate a high central surface brightness for the underlying partially extended MIR emission (Figure \ref{psf_modelling_example}). However, with careful examination, extended structure is visible, particularly at 5.6 \mics\ and 10 \mics. North of the nucleus, at a PA$\approx -12^{\circ}$, a broad tongue of emission can be discerned with a maximum extent of $1\farcs7$. A counter-feature to the south may also be visible, though marginally in the current images. The northern structure is cospatial with bright emission in [O III]$\lambda 5007$ seen in high-resolution HST narrow-band images, as well as a filament of molecular gas seen in the CO map in Figure \ref{ngc3227}. In the [Si VI] map shown in the Figure, the northern feature does not distinguish itself strongly over the fairly smooth and isotropic shape of the NLR. 

The extended star-forming disk of NGC\,3227 is clearly seen in our images, even though its lower surface brightness compared to the nucleus makes it hard to see in the current stretch of Figure \ref{ngc3227}. A detailed study of the various galaxy components detectable in the images will be presented in later work, but here we highlight a few bright knots that stand out in MIR emission of the disk. A prominent knot is visible in Figure \ref{ngc3227} 2'' south and 4'' west of the nucleus. It lines up well with a knot of molecular gas seen in CO(2-1), though no optical dust feature or NIR emission line knot (in [Si VI] or Br$\gamma$) can be seen. These knots are likely to be compact regions of strong star-formation, perhaps large star clusters a few 100 pcs from the nucleus.

\subsubsection{ESO 428-G14}

\begin{figure*}
	\centering
	\includegraphics[width=\textwidth]{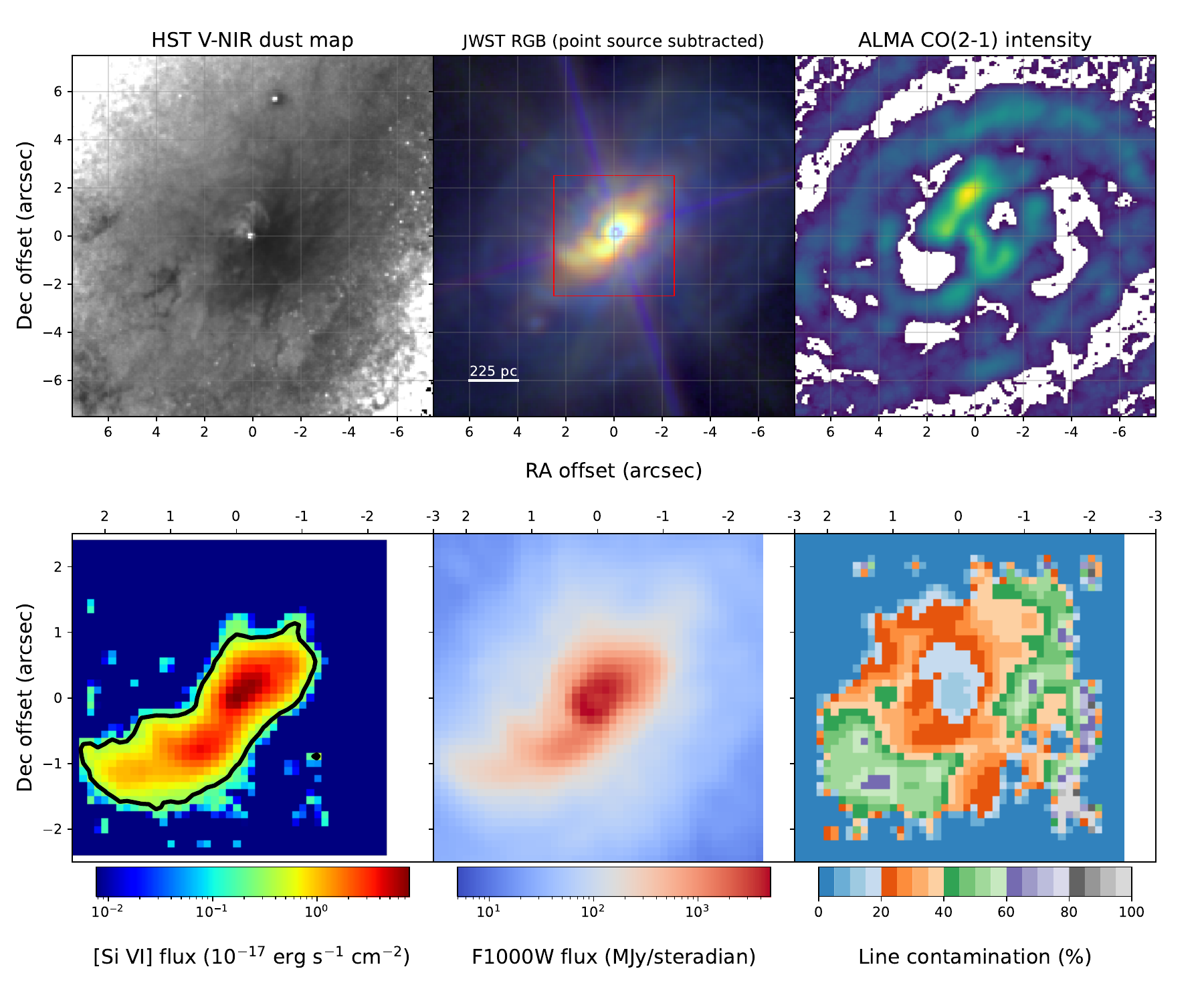}	
	\caption{Context figures for ESO 428-G14. See the caption for Figure \ref{ngc4388} for explanations for each of the panels. The only difference is the {\it HST} structure map in the upper left, used to reveal the dust lanes. Unlike the other targets, the structure map is produced using a single optical image. 
	}
 	\label{eso428}
\end{figure*}

ESO 428-G14 (Figure \ref{eso428}) is a bulge-dominated spiral galaxy, classified as S0 from HyperLEDA. Based on spectral fitting of the X-ray emission from the Compton-thick AGN, \citet{levenson06} report a minimum 2-10 keV luminosity of $\gtrsim 3.5\times10^{41}$ \ergs. Even if the luminosity was higher by several factors, ESO 428-G14 is the least luminous AGN in our sample. Despite this, it has a significant NLR with a high surface brightness. Previous studies have suggested that the emission lines may be partly powered by fast shocks driven through the NLR by the radio jet in this system \citep{riffel06, may18, fabbiano18}.

There is no existing HST imaging in the NIR for ESO 428-G14, so we do not have a resultant V-NIR colour map in Figure \ref{eso428}. Instead, following \citet{pogge02}, we create a ``structure map'' from the F814W image of the galaxy, designed to bring out the faint dust features around the nucleus. The tendrils of dust absorption trace the tightly wound circumnuclear spiral \citep[first reported in][]{prieto14}, also evident in the CO map. The circumnuclear disk is seen in the short wavelength JWST images from our programme, but the brightest extended emission is in a set of high surface brightness features that run SE-NW through the nucleus, aligned with the NLR, ionisation cone, and radio jet. Our emission line contamination map shows that a large fraction of the SE MIR emission, and some of the NW emission $>1$ arcsecond from the nucleus, is affected by NLR emission lines.

ESO 428-G14 is the subject of an in-depth GATOS-led JWST imaging study \citep{haidar24}. We find that, after correcting for the contribution from lines, the extended dust continuum emission closely matches the structure of the asymmetric radio jet. We suggest that ESO 428-G14 may show some of the first evidence for detectable dust heating by outflow-mediated radiative shocks. More details, as well as a complete discussion of the circumnuclear region in this galaxy, can be found in \citet{haidar24}.

\subsubsection{NGC\,3081} 

\begin{figure*}
	\centering
	\includegraphics[width=\textwidth]{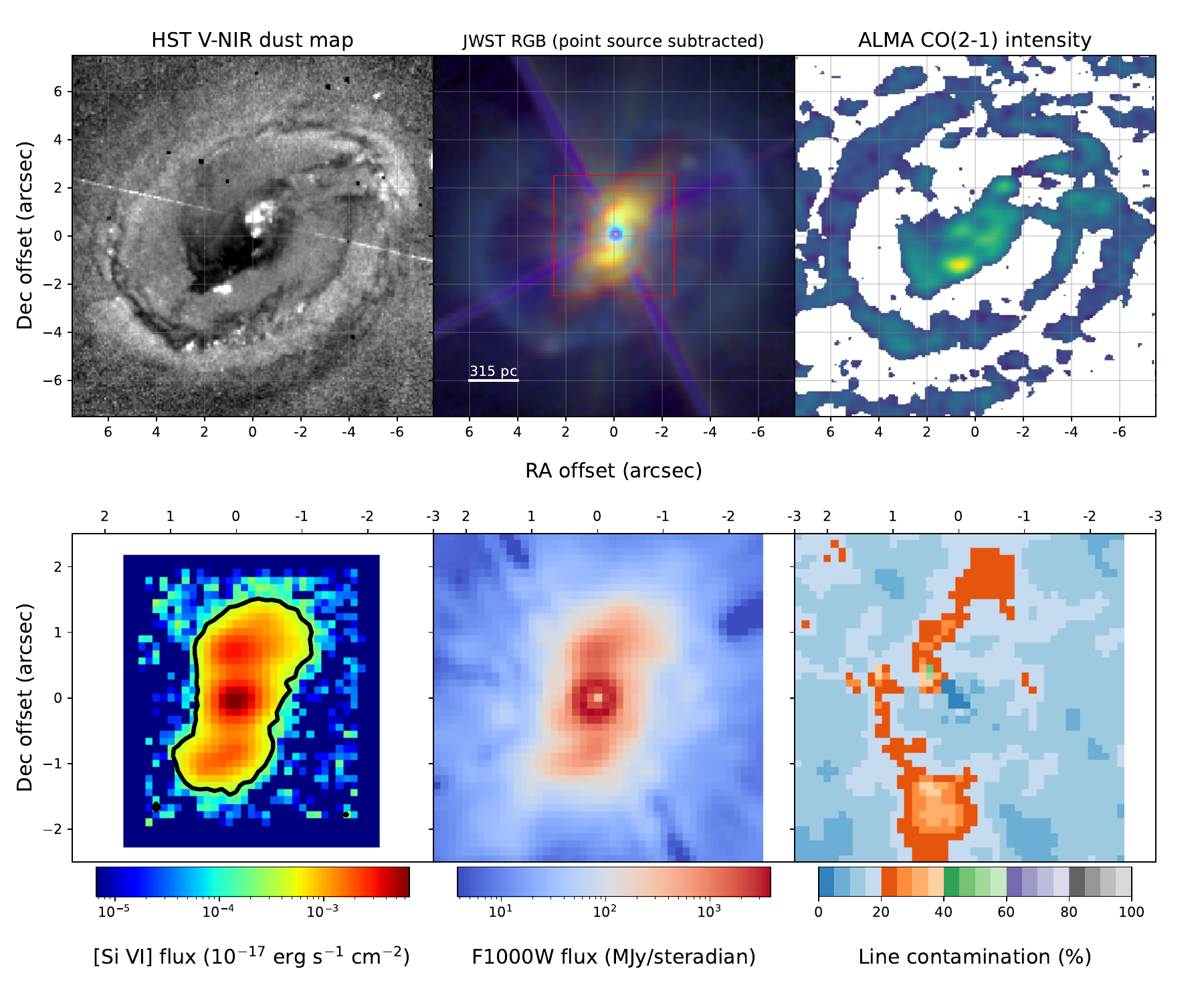}
	\caption{Context figures for NGC\,3081. See the caption for Figure \ref{ngc4388} for explanations for each of the panels.
	}
	\label{ngc3081}
\end{figure*}

A galaxy with multiple spectacular resonance rings, NGC\,3081 (Figure \ref{ngc3081}) is a textbook case for the study of barred galaxy dynamics \citep{buta98}. In the top left panel of Figure \ref{ngc3081}, we see the circumnuclear ring, 15'' across its major axis, from which two spiral arms, delineated by dust, flow towards the nucleus. The optical-NIR colour map also shows star-forming regions in the nuclear ring, and bright emission line knots around the nucleus from NLR gas. The dusty region surrounding the inner 2'' of the nucleus is mirrored in the CO(2-1) map as a region with patchy molecular gas emission. The ring and spiral arms are also traced by molecular gas.

The NLR traced by optical emission lines is very asymmetric, and brighter north of the nucleus \citep{ferruit00}. However, from NIR spectroscopic observations, it is clear that the NLR is intrinsically much more symmetric (e.g., see the [Si VI] map in the lower left panel of Figure \ref{ngc3081}), and its optical appearance is heavily modulated by dust extinction. Its structure is ``S''-shaped, with a central beam roughly 2'' long that is oriented due North-South, from which more diffuse emission bends off toward the west and east at the northern and southern extent respectively.

A compact outflow has been identified in this AGN from HST slitless spectroscopy \citep{ruiz05} and ground-based optical IFU spectroscopy \citep{schnorr-muller16}. Strong blueshifts and line FWHM $> 500$ \kms\ are associated with the inner 2'' around the nucleus, corresponding roughly to the North-South beam of the NLR and the high surface-brightness central knot seen in [Si VI]. The more diffuse NLR extensions, further to the north and to the south, only show mild velocities and linewidths, with a general redshift to the north and blueshift to the south. This implies an outflow that is oriented so that the southern part is angled towards the observer. However, the kinematics of the outflow are complex and cannot be characterised by a simple model, such as an evacuated bicone \citep{fischer13}.  

In the JWST images, we clearly see the circumnuclear ring in the MIR, appearing with a blue colour consistent with the flat SED from dust heated by localised star-formation (Section \ref{mir_colours}). The brightest emission is in the central few arcseconds, where a bipolar symmetric structure with redder colours can be seen. The shape and orientation of the bipolar structure mirrors the NLR so well that one may expect a strong contribution from emission lines in the MIR emission, after our cautionary notes from Section \ref{emlinecont}. However, both our contamination map, and, more directly, the existing MRS spectroscopy of this AGN shows very low contamination from emission lines in this structure. It is primarily dust continuum emission that traces the extent of the illumination pattern of the central source. As the most intrinsically luminous AGN in our sample, it is likely that the extended MIR continuum emission in NGC\,3081 is mostly due to {\it in-situ} NLR dust heated by the AGN.

\subsubsection{NGC\,7172}

\begin{figure*}
	\centering
	\includegraphics[width=\textwidth]{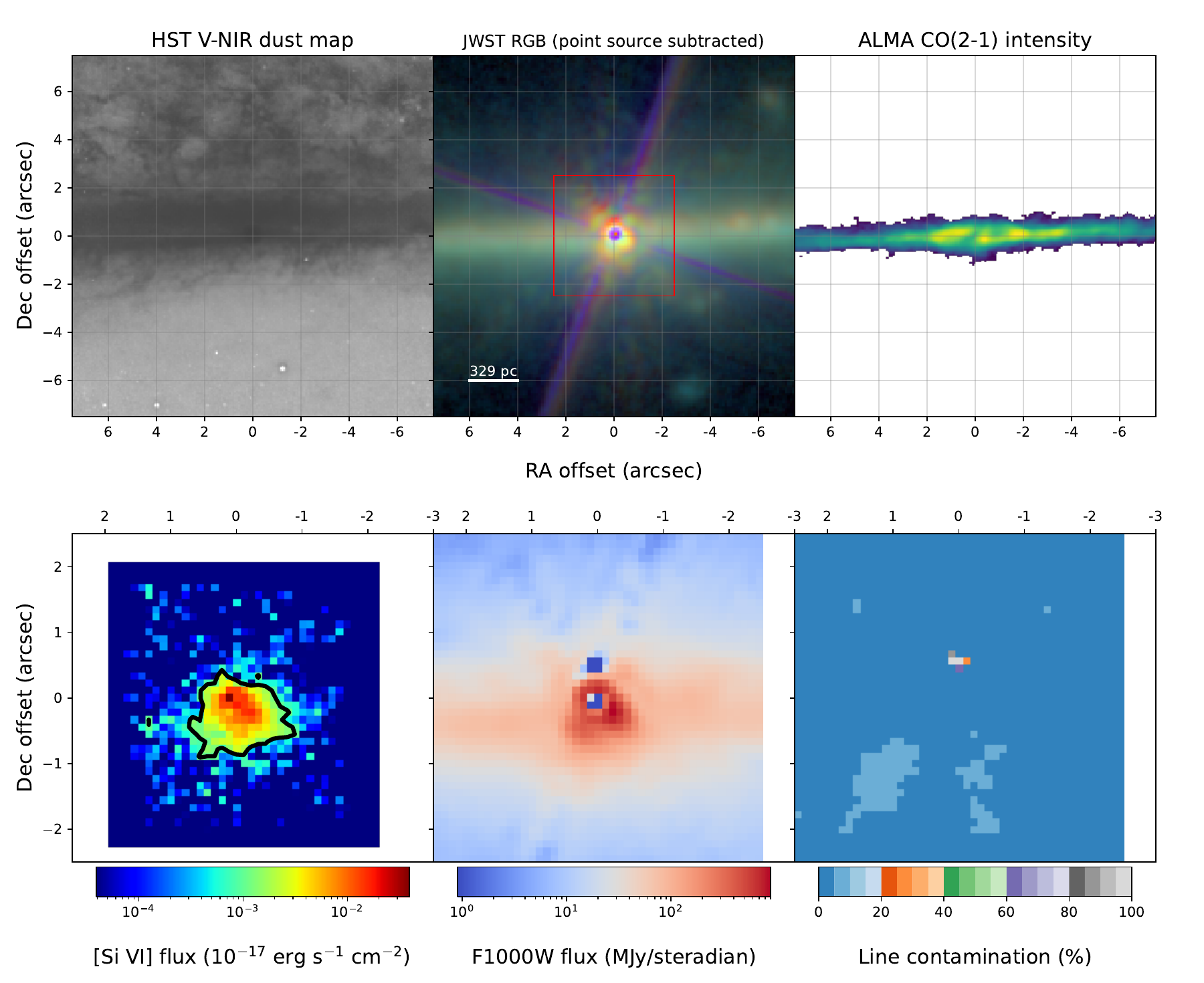}
	\caption{Context figures for NGC\,7172. See the caption for Figure \ref{ngc4388} for explanations for each of the panels.
	}
	\label{ngc7172}
\end{figure*}

NGC\,7172 (Figure \ref{ngc7172}) is an almost edge-on spiral galaxy \citep[PA\,$\sim 88^{\circ}$;][]{alonso2023}, with a prominent dust lane covering the nucleus \citep[A$_{V} > 3$; e.g.][]{smajic12} and a circumnuclear star-forming ring of $\sim$1-1.4\,kpc in diameter \citep{alonso2023}.
As one of the first systems shown to host a heavily-obscured Seyfert nucleus \citep{sharples84}, the galaxy has been the focus of several studies that have examined the properties of its ionised and molecular gas \citep{smajic12,thomas17,davies20,alonso23,hermosamunoz24}. 
It exhibits a two-sided ionisation cone in high-excitation lines, oriented N-S, and extending out to $\sim 900$ pc from the centre. In the optical, the northern cone is strongly extinguished by the galaxy's dust lane. NIR spectroscopy reveals broad Pa$\alpha$ lines, implying that the illumination of the AGN may be close to the line of sight, despite the apparent optical Type\,2 nature of the galaxy. This is further supported by the rapid variability of the nucleus in the hard X-rays \citep[e.g.][]{guainazzi98} and the relatively strong emission from hot dust (temperatures $\sim 1500$ K) in the IR \citep[][see also Figure \ref{all_sedfig2}]{burtscher15}.

 \cite{davies20} found velocities up to $\sim 400$ \kms\ for the [O III]$\lambda5007$ line at a deprojected distance of $\sim 160$ pc, from which they derived a mass outflow rate for the ionised gas of $0.005$ M$_{\odot}$ yr$^{-1}$, a low value compared with other Seyfert galaxies of similar luminosities from their sample. \cite{alonso2023} also detected the inner part of the southern region of the outflow with VLT/SINFONI using the [Si VI]$\lambda 1.96$ \mics\ emission line, extending out to $\approx 200$ pc. Additionally, they detected hints of a molecular outflow, also noticed by \citet{stone16}, seen with the CO(3-2) emission in the direction of the ring, moving at an average velocity of 46 \kms, carrying mass out at a rate of $40$ M$_{\odot}$ yr$^{-1}$, significantly larger than the measurements of the ionised gas, probably due to a stronger coupling with the ISM \citep{alonso2023}.
 
 In our MIRI images, we clearly detect the edge-on galaxy disk, showing the same structure and thickness as the molecular gas disk (central and right upper panels of Figure \ref{ngc7172}). The central MIR source is very prominent, but, under close examination, our point source-subtracted images show a distinct extension to the SW which overlaps with the asymmetric outflowing part of the NLR seen in [Si VI] (e.g., lower left panel of Figure \ref{ngc7172}). Direct measurements from existing MRS spectra show low levels of emission line contamination in the outflow, which is consistent with our estimated contamination using NIR spectra \citep[Section \ref{emlinecont} and][]{campbell25}. In the same vein as ESO 428-G14, the high degree of spatial correlation between the extended outflow and this dust continuum feature suggests that we are observing warm dust within this wind. Future work will explore this association in more detail.
 
 Within 5'' of the nucleus, and out of the plane of the galaxy disk, we also observe diffuse clumps of MIR continuum emission that lie within the illumination cone of the central nucleus. Particularly visible to the SW, these clumps have blue MIR colours suggestive of star-forming regions, and may trace star formation entrained within a broader slow wind, or initiated by it. Interested readers can find more details in \citet{hermosamunoz24}.

\subsubsection{NGC\,2992}

\begin{figure*}
	\centering
	\includegraphics[width=\textwidth]{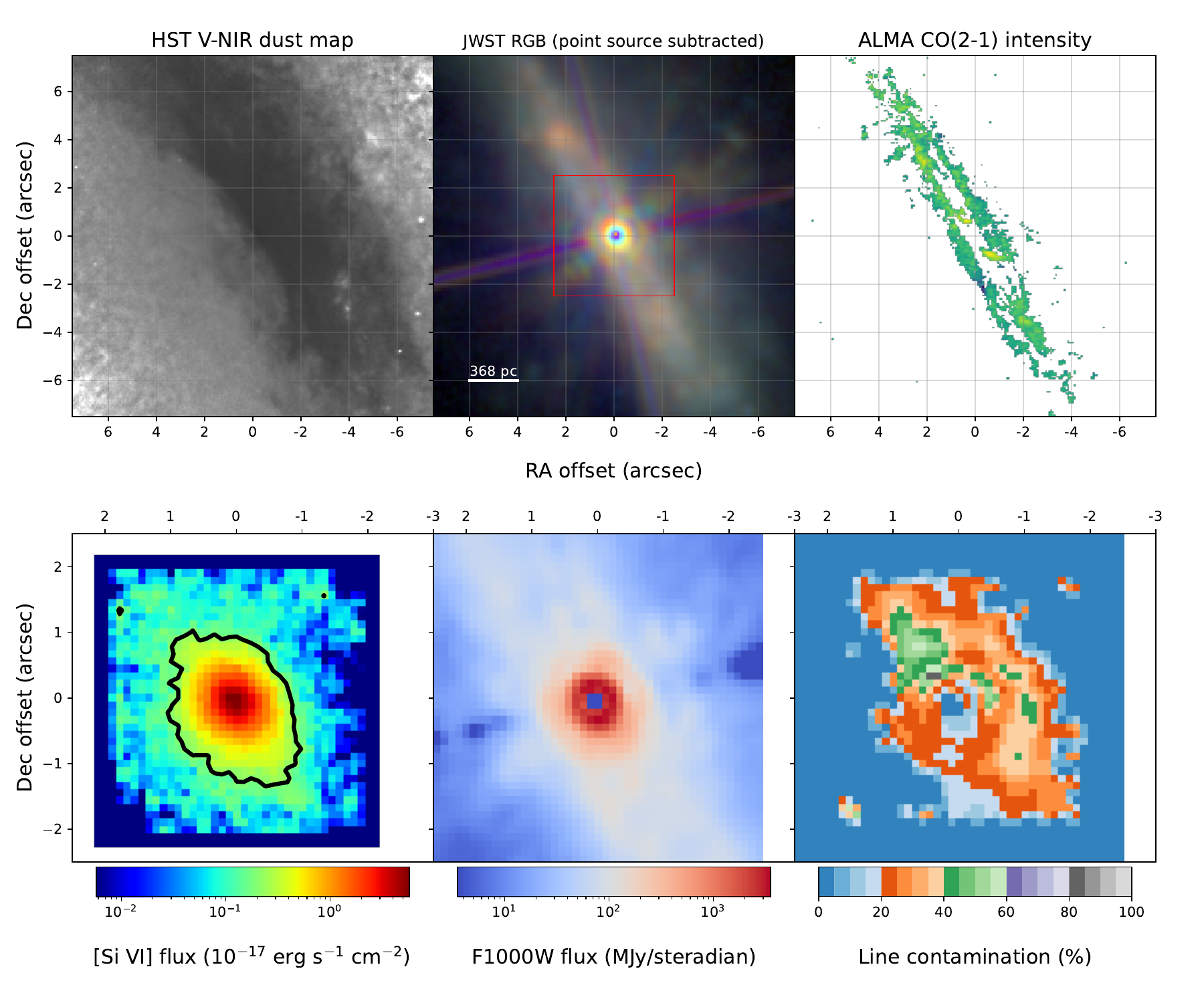}
	\caption{Context figures for NGC\,2992. See the caption for Figure \ref{ngc4388} for explanations for each of the panels.
	} 
	\label{ngc2992}
\end{figure*}

NGC\,2992 (Figure \ref{ngc2992}) is an inclined spiral galaxy, part of the interacting system Arp 245, which also includes NGC\,2993 and a tidal dwarf. The properties of the host galaxy indicate that it going through the early stages of an interaction \citep{garciabernete15}. Although classified as a Seyfert 1.9 in the optical, its AGN type has shifted over time \citep{trippe08}, consistent with a Type\,1 nucleus of variable luminosity observed through a dust lane with a moderate extinction of $A_{V} \approx 3$. This is additionally supported by the large nuclear variations in the X-rays \citep{gilli00} and in the IR \citep{glass97}.  

A thin host galaxy disk and central ring is clearly seen in CO(2-1), giving a galaxy inclination of $\sim80^{\circ}$ \citep[Figure \ref{ngc2992}, consistent with][]{zanchettin23}. In the UVO, a prominent thick dust lane is present with a PA $\sim 30^{\circ}$ \citep{colina87}. In ground-based N-band imaging (at 11.2 \mics), faint extended emission is seen cospatial with the disk, extending out to approximately $\sim 3$ kpc \citep{garciabernete15}. These authors reported the presence of PAH emission in the disk, relating it to star-formation heating. Interestingly, they found that PAH features have been destroyed in the inner $\sim50$ pc, likely by the influence of the AGN.

The extended biconical NLR extends out to $\sim 5$ kpc, with the southeastern cone located in front of the galaxy disk and the base of the northwestern outflow located behind it \citep{colina87,garcia-lorenzo01,veilleux01, zanchettin23}. The optical opening angle of the cone is large \citep[$\sim 120^{\circ}$][]{veilleux01, garcia-lorenzo01, zanchettin23}, but this is at least partly shaped by the dust lane because the NLR in the NIR does not display any bipolarity at scales of $\sim 3$''. Despite a vigorous level of star-formation in the central regions of the host galaxy, emission line diagnostics indicate the NLR is primarily ionised by the AGN, either by nuclear radiation or fast radiative shocks \citep[e.g.,][]{allen99, garcia-lorenzo01}.

An outflow is well-established in this AGN, both on nuclear scales in the X-rays \citep{marinucci18} and on kpc-scales in ionised gas. The detailed structure and kinematics of the ionised outflow has been difficult to pin down because of the important contribution of the galaxy disk in projection towards the NLR. In a recent study built upon a VLT/MUSE IFU dataset, guided by ALMA CO spectroscopy, \citet{zanchettin23} demonstrated that, within the inner 600 pc, there is a fast, compact outflow with maximal velocities of $\approx 1000$ \kms, blue-shifted towards the SE and red-shifted towards the NW. At larger radii, the outflow quickly loses definition and the kinematics appear to be rotational. A number of gas clouds are seen in CO(2-1) perpendicular to the galaxy disk, which may be carried out by the interaction between the outflow and the galaxy disk at the edges of the NLR \citep{zanchettin23}.

Our MIRIM images clearly show the inclined galaxy disk, as one may expect from the prominence of its emission in ground-based MIR studies \citep{garciabernete15}. We can easily delineate the dust that corresponds to the molecular gas ring seen in the CO map. Similarly to  NGC\,4388 and NGC\,7172, the other highly inclined galaxies in our sample, the galaxy disk in the MIR dust continuum is comparable in thickness to the CO disk, but thinner than the dust lanes that are traced by the HST imaging. The F1000W-F1500W colour of the galaxy disk is fairly red, paralleling that of NGC\,7172. This may indicate a non-negligible amount of line-of-sight MIR obscuration in these massive gas-rich disks.

The extended polar emission in NGC\,2992 is not particularly prominent, and we do not see any clear emission that suggests a bipolar structure like the optical ionisation cones. We detect a number of faint clumps perpendicular to the galaxy disk, some of which line up with the faint CO knots reported in \citet{zanchettin23}. These may be the dusty counterparts of the cool clouds raised by the central outflow. However, despite the low levels of emission line contamination in the JWST images, we do not detect any prominent polar emission in this object, either from the known outflow or cospatial with the large radio jet found in this system \citep{ulvestad84, haidar26}.

\subsubsection{NGC\,5728}

\begin{figure*}
	\centering
	\includegraphics[width=\textwidth]{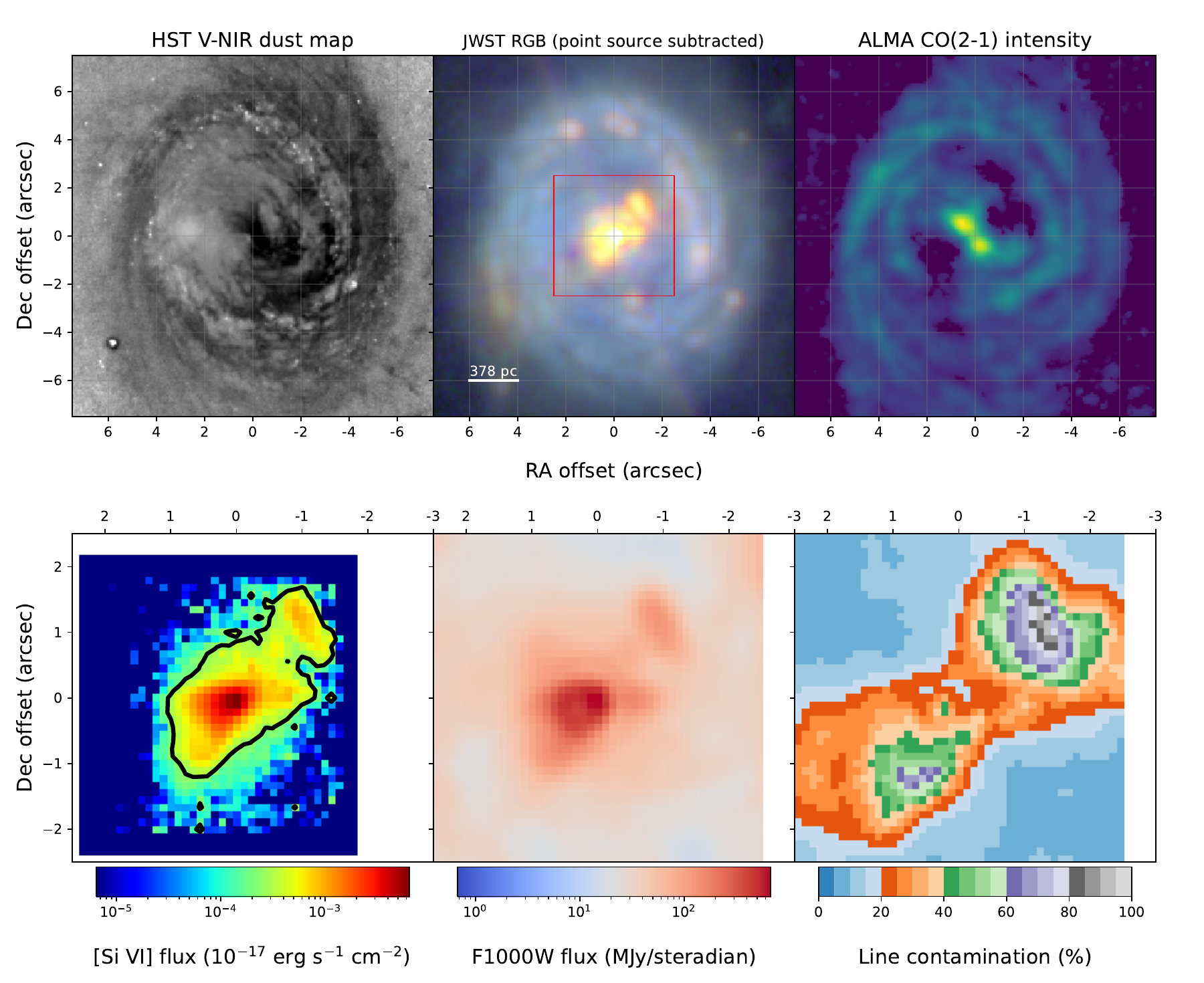}
	\caption{Context figures for NGC\,5728. See the caption for Figure \ref{ngc4388} for explanations for each of the panels.
	} 
	\label{ngc5728}
\end{figure*}

NGC\,5728 (Figure \ref{ngc5728}) is a nearby SAB(r)a disk galaxy with a Compton thick Seyfert 2 nucleus \citep{phillips83}. The host galaxy has a large-scale stellar bar, $\approx 11$ kpc across \citep{rubin80} and a prominent circumnuclear star-forming ring $\approx 800$ pc in radius \citep{schommer88, shaw93}.

The NLR in this system has been extensively studied, and is considered one of the best cases of a biconical illumination nebula ionised by an AGN \citep{wilson93, son09, durre18, shimizu19, durre19}.
The cone is fairly narrow, with an extent of $\approx 2$ kpc, an opening angle of $\approx 60^{\circ}$, and a PA$\approx 118^{\circ}$. The SE side of the ionisation cone is expected to lie in projection above the disk of the galaxy \citep[e.g.,][]{son09}. A radio jet has been mapped with high-resolution VLA imaging, aligned with the PA of the ionisation structure, showing bright narrow structure towards the SE and a more diffuse lobe-like counterstructure to the NW \citep{schommer88}.

Kinematic studies based on a number of emission lines indicate a fast outflow within the inner 250 pc of the nucleus.  
Its orientation and derived geometry implies that it intersects the plane of the circumnuclear disk delineated by the star-forming ring \citep{shimizu19, davies24}.

The central region of NGC\,5728 was the focus of a detailed JWST MIRI/MRS spectroscopic study that explored the interaction between the disk, the ionisation cone, and the outflow \citep{davies24}. Prominent holes in the distribution of low-excitation CO in the circumnuclear disk (e.g., see Figure \ref{ngc5728}) are co-spatial with the outflow, indicating that its impact alters the local properties of molecular gas. These regions contain a larger proportion of warm molecular gas, traced by the MIR H$_{2}$ rotation lines. \citet{davies24} suggest that shocks imparted by the outflow on the molecular gas are primarily responsible for heating it, and also for accelerating the gas at scales $< 50$ pc.   

The circumnuclear disk and its associated star-forming regions are the most prominent features in the MIRIM images of NGC\,5728. Within the central 4 arcseconds, we can discern an extended structure that is aligned along the PA of the ionisation cone and with a higher surface brightness than the surrounding disk. This emission shares a lot of structure with the ionised gas NLR traced by the [Si VI] line (lower panels of Figure \ref{ngc5728}). Our line contamination analysis shows that a significant part of the MIR structure in the images, particularly at 10 \mics, can be attributed to emission lines in the MIRI filter bandpass. Nevertheless, after correcting for emission lines using the MRS spectra available for this AGN, we find clear extended dust continuum emission that is cospatial with the NLR \citep{haidar26}.   

Particularly for the star-forming disk, there is a strong correspondence between the structures seen in CO and in the MIR dust continuum, many of which are also seen in absorption from the \textit{HST} V-NIR dust map (upper panels of Figure \ref{ngc5728}). In the central few arcseconds, the CO map shows a prominent pair of knots on either side of the nucleus, oriented along a PA$\approx 45^{\circ}$ and coincident with a very dark dust absorption in the V-NIR colour map. From previous work, this is known to be an edge-on dense molecular gas disk aligned with the axis of the AGN's accretion disk, and almost perpendicular to the plane of the circumnuclear disk \citep{shimizu19}. The strong silicate absorption seen towards the nucleus of NGC\,5728 very likely originates in this disk. Despite its prominence at other wavelengths, it does not appear in our MIR dust images, probably because most of the dense gas disk is cold and does not emit strongly in the MIR.

\subsubsection{NGC\,5135} \label{ngc5135-text}

\begin{figure*}
	\centering
	\includegraphics[width=\textwidth]{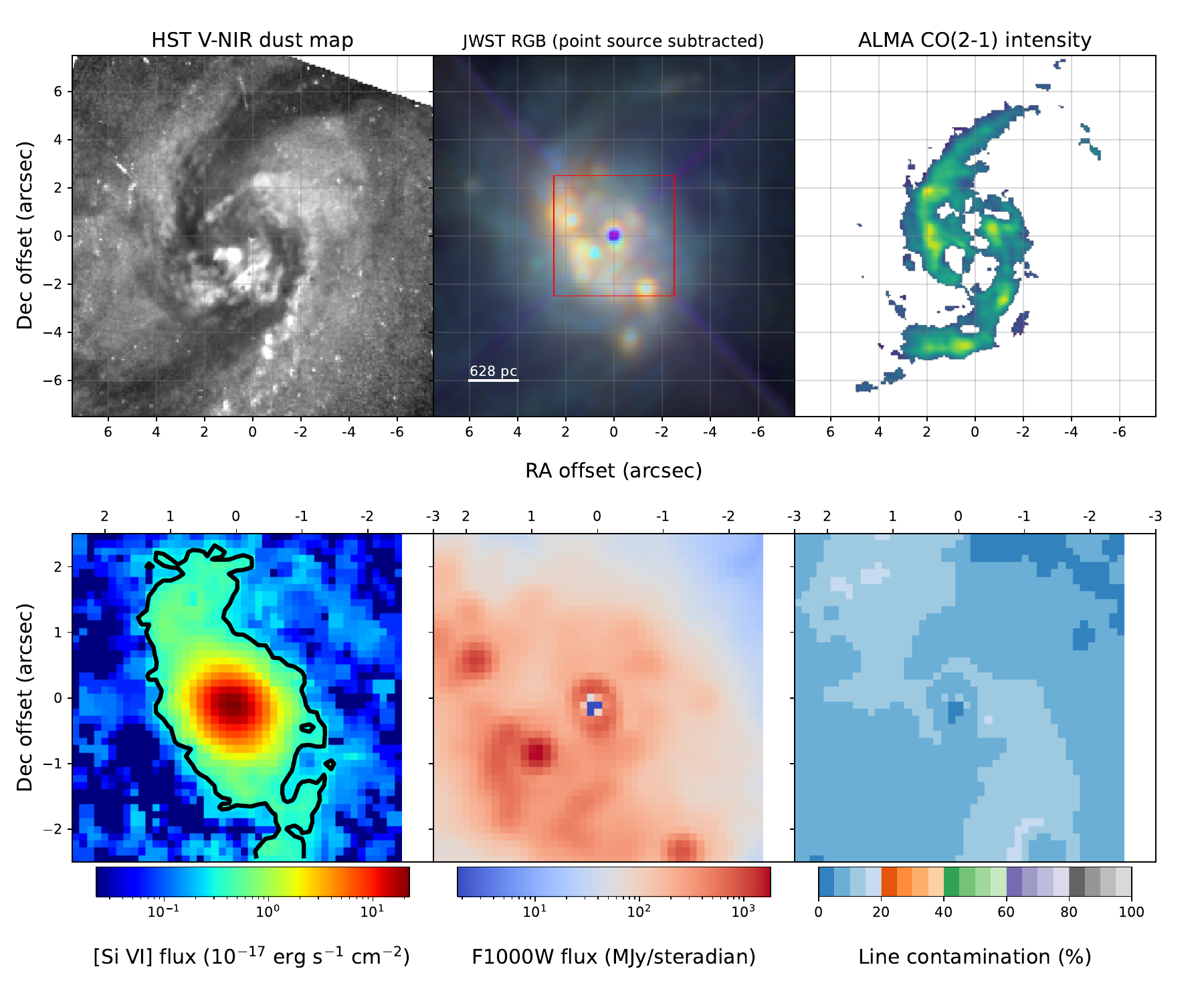}
	\caption{Context figures for NGC\,5135. See the caption for Figure \ref{ngc4388} for explanations for each of the panels.
	}
    \label{ngc5135}
\end{figure*}

NGC\,5135 (Figure \ref{ngc5135}) is a face-on barred spiral hosting a Luminous InfraRed Galaxy \citep[LIRG, $L_{\textrm{IR}} = 10^{11.3} L_{\odot}$;][]{pereira-santaella11} and a Compton thick (CT) AGN \citep[][]{levenson04, ricci21}. The central few kpc feature a large circumnuclear disk studded with several intense star-forming regions \citep[e.g][]{alonso-herrero06} that sustain the central starburst in the galaxy. The disk is bounded by grand-design spiral arms to the north and south. These connect to the leading edges of the large-scale bar that is oriented NW to SE.

The AGN-ionised NLR in NGC\,5135 takes the form of an ionisation cone, detected most clearly in high-excitation lines such as [Si VI]$\lambda 1.96$ \mics\ \citep{bedregal09}, extending NE and SW $\sim$ 600 pc from the nucleus. There is no clear indication of a large-scale outflow, with the line profile remaining narrow (FWHM $< 160$ \kms) over the extended NLR. However, towards the nucleus, there is a spatially-compact region (size $< 200$ pc) that shows a strong blue-shifted wing (FWHM $\approx 430$ \kms), indicating a localised small-scale outflow. 

In the MIRI images (see the middle panel of the top row in Figure \ref{ngc5135}), the structural definition of the circumnuclear disk becomes clearer, especially in comparison to the HST images and dust map (left panel of corresponding Figure). The various dust-obscured star-forming regions are revealed as distinct knots, most of which line up with the positions of HII regions seen in Pa$\alpha$ \citep{alonso-herrero06}. Some of the brightest knots in the MIRI images, such as those due east of the nucleus, are invisible in the optical HST image because of intervening dust, though they have counterparts in the molecular gas distribution (right panel of the top row in Figure \ref{ngc5135})

The ALMA and MIRI images show broad similarities, but some important differences as well. In contrast to the CO map, the eastern half of the disk is brighter relative to the western half in the MIRI images, and the disk itself has a higher contrast compared to the large-scale spiral arms. If the extended MIRI emission is primarily dust continuum, as we expect in images of NGC\,5135, this may be an indication of a high and variable efficiency of star-formation in this circumnuclear starburst. 

 Most peculiar however, is a compact region to the south-east of the nucleus, which is particularly bright at 10 $\mu$m as seen in the F1000W-1500W color map in Fig. 10. This would suggest silicate emission from a star-forming region, which is typically only present in Type\,I AGN while star-forming regions show moderate silicate absorption \citep[e.g.][]{gallimore10, garcia-bernete17, garcia-bernete22}. 

In addition to the regions discussed above, X-ray emission is detected 0.9 kpc south of the AGN \citep[e.g.][]{colina12}, coincident with strong [Fe II] (1.64 $\mu$m) and warm H$_{2}$ (2.12 $\mu$m) \citep{bedregal09}, attributed to supernova driven shocks. This region is kinematically independent, with a bi-conical shaped velocity gradient centered on the [Fe II] peak with velocities of $\sim$ 90 kms$^{-1}$. There is no clear counterpart in the MIRI images, however there is a mid-IR bright region nearby.

\subsection{Spatial correlations between dust continuum and the NLR}

\begin{figure*}
	\centering
	\includegraphics[width=\textwidth]{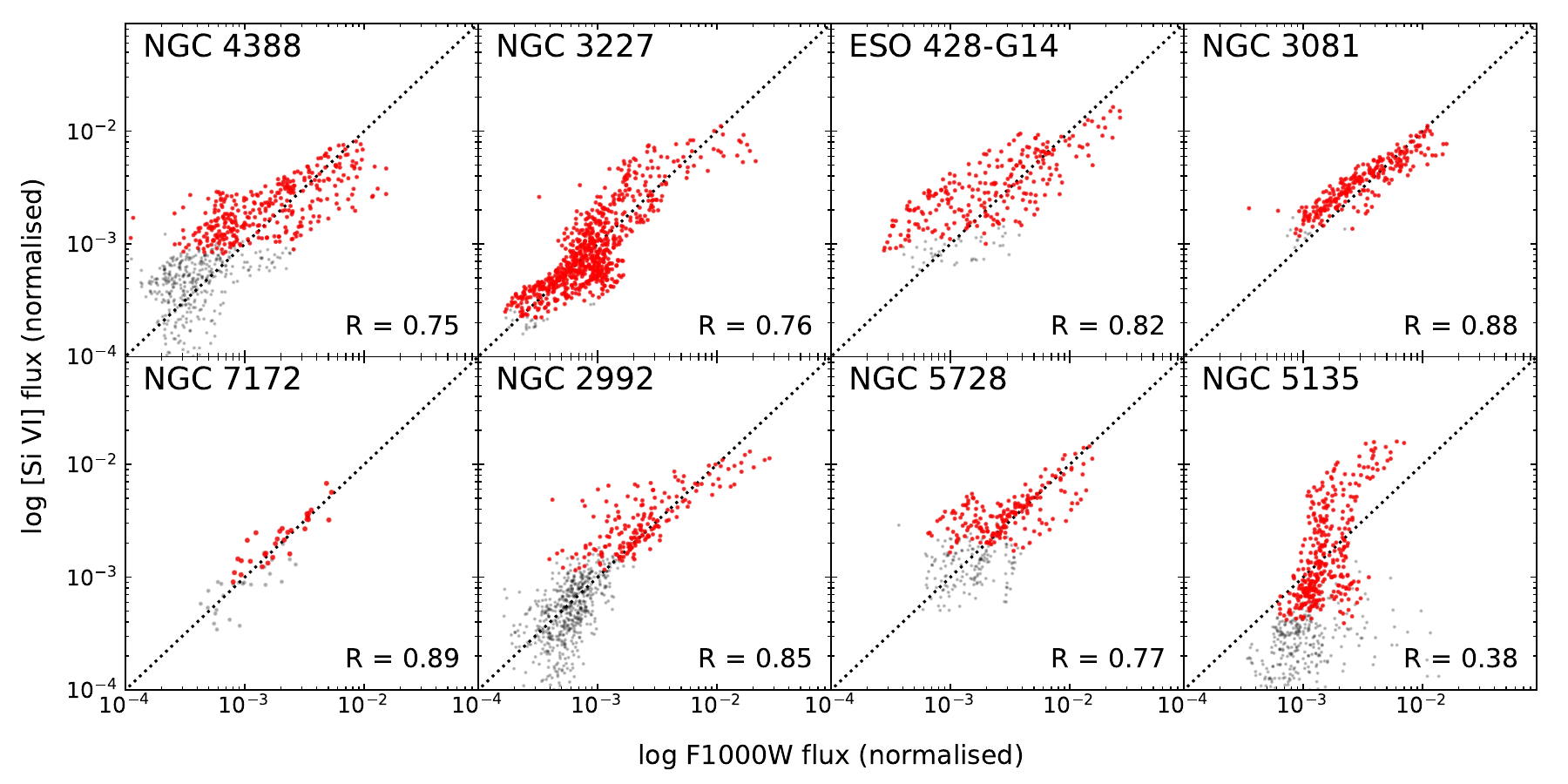}
	\caption{A comparison of normalised pixel-by-pixel fluxes of the [Si VI]$\lambda 1.96$ \mics\ emission line (from VLT/SINFONI) and the 10 \mics\ dust continuum (from \jwst\ F1000W images after PSF subtraction and correction for line contamination) for all the Seyferts in our sample. Red/grey points indicate pixels within/outside the narrow-line region (NLR). The dotted line shows the 1:1 line. The value of the Pearson correlation coefficient (R) calculated using all the pixels is given in the right bottom corner of each panel. All pixels within $0\farcs66$ of the nucleus have been removed from this analysis.
	}
	\label{spatial_correlations}
\end{figure*}

An important visual result from our multiwavelength analysis of the Seyferts is the strong spatial correpondance between the extended F1000W emission within the inner few arcseconds, and the [Si VI]$\lambda 1.96$ \mics\ maps (left and middle panels in the lower row of Figures \ref{ngc4388} -- \ref{ngc5135}). This is clear in most of the galaxies with bright polar MIR emission (NGC\,4388, ESO 428-G14, NGC\,3081, NGC\,7172, NGC\,5728), but less well-defined in those with circularly-symmetric emission line regions (NGC\,3227 and NGC\,2992) and in the starburst-dominated NGC\,5135.

The spatial association between dust continuum and emission lines has been explored in earier GATOS papers. Using this imaging dataset, \citet{haidar26} highlighted a strong visual correpondence between [Si VI] and F1000W emission, after correcting the \jwst\ images for emission-line contamination. Their results mirror the impressions outlined above. In contrast, \citet{lopez25} reported a rather weak spatial correlation between dust continuum emission and MIR emission lines (including high-excitation lines such as [Fe VII]$\lambda 9.52$ \mics) in a sample of similar Seyfert galaxies, including three that overlap with our sample. Their work was based on MIRI/MRS IF spectroscopy, which allows a clean separation of continuum and line emission, but suffers from worse spatial resolution. 

We undertook a spatial correlation analysis, similar to that of \citet{lopez25}, to numerically assess the relationships between dust continuum and NLR line emission. For this we use the [Si VI] maps (Section \ref{si6}) and the PSF-subtracted F1000W images after subtracting the best estimate of the emission-line contamination (Section \ref{emlinecont} and right lower panels in Figures \ref{ngc4388} -- \ref{ngc5135}). The contamination for NGC\,3081, NGC\,7172 and NGC\,5728 was estimated from the same MIRI/MRS spectra that were used in \citet{lopez25}.

After matching astrometry and spatial sampling, we masked both images to exclude pixels that had no detectable [Si VI] flux, as well as all pixels within a radius of $0\farcs33$ from the nucleus, to minimise the effects of PSF subtraction residuals. We then normalised both maps so that the fluxes could be compared on the same numerical scale, by dividing by the total flux in the unmasked pixels. This rescaling does not affect any spatial correlation results.

We calculated Pearson correlation coefficients ($R$)between the scaled fluxes of [Si VI] and 10 \mics\ continuum for all unmasked pixels. The flux comparisons and the correlation coefficients are shown in Figure \ref{spatial_correlations}, where we distinguish between pixels that lie within (red points) and outside (grey points) the NLR masks defined in Section \ref{cmap_results}.

Except for NGC\,5135, which shows no evidence for NLR-aligned dust emission (Section \ref{ngc5135-text}), the correlation coefficients are high ($R > 0.75$). The coefficients change a little depending on whether we consider only pixels within the NLR or include all pixels in the masked maps, but are always high. This confirms the visual impressions reported in \citet{haidar26}, but remains in tension with the average coefficient values of $<0.5$ found by \citet{lopez25}.

A closer examination of the correlations in Figure \ref{spatial_correlations} suggests a diversity of pixel flux relationships between the two tracers. Even in galaxies with clear visual connections between [Si VI] and the dust continuum, the correlation coefficients vary between $0.75$ for NGC\,4388 and $0.89$ for NGC\,7172, indicating that there is substantial scatter in these relationships. Some of this scatter is likely driven by the prominence of dust emission heated by star-formation that underlies the AGN-heated dust. For example, in the case of NGC\,5728, which has a bright star-forming circumnuclear disk, we can discern, among the red points (within the NLR), a tight sequence associated with the emission-line region, and a sequence with more scatter that likely comes from pixels with substantial disk emission. The regions that lie outside the NLR all show a larger scatter in this object.

We suggest that the different levels of spatial correlation reported in earlier GATOS works is driven by differences in the spatial resolution between the MIRI imager and MRS, as well as the effects of the unsubtracted nuclear point source in the work of \citet{lopez25}\footnote{PSF modelling and subtraction for the MIRI/MRS is very difficult, and is only now reaching a level of maturity \citep[e.g.,][]{gonzalez-martin25}}. The lower resolution of the spectroscopic study likely averaged over regions that contained disk continuum emission, which is not strongly correlated with high-ionisation emission lines, and AGN-heated continuum emission, leading to a lower measured value of the spatial correlation coefficient.
Based on our analysis using the higher resolution PSF-subtracted MIRIM images, we conclude that there is a substantial connection between the AGN-heated dust and high-ionisation emission lines, in line with the results of \citet{haidar26}. 

\subsection{Polar dusty outflows?}

While AGN-heated dust emission in the polar direction is found in the majority of objects in this study, can we conclude whether this emission originates from a dusty wind accelerated from the torus? There is a pronounced similarity between the polar MIR dust structures around the nuclei, and the structures traced by high-ionisation emission lines that arise in the radiation field of the AGN. These congruences indicate that the likely heating source of the warm dust is nuclear radiation, and that the dust follows the gaseous distribution that also determines the appearance of the NLR. 

This is consistent with recent theoretical understanding which indicates that the radiation pressure-driven wind may not be able to sustain itself beyond $\sim 10$ pc \citep[e.g.,][]{venanzi20, williamson20}, because of the rapid decrease in the optical depth of the dust, as well as interactions with the extended ISM of the host galaxy. Therefore, it is unlikely that the AGN-heated dust that we see in our images is part of a continuous outflow from the nucleus. However, as demonstrated in \citet{haidar24} for ESO 428-G14, there is evidence that the highest surface brightness dust emission is cospatial with regions that also show kinematic signatures of fast gaseous outflows. At present, our best hypothesis is that the shocks that accompany the outflows from the AGN are able to heat dust without completely destroying it \citep{haidar26}. It is likely that the dust emission that we are seeing in the NLRs are a combination of such shock-heated dust and more pervasive, but lower temperature, dust heated by nuclear radiation. 

\section{Conclusions}

The Dust in the Wind programme (PID 2064) gives us a detailed look at the MIR structures in the central few kpc of a diverse set of local Seyfert galaxies, showcasing the ground-breaking capabilities of \jwst\ MIRIM imaging. This paper serves as a reference for the reduction, data processing, and analysis techniques used to characterise the nuclear and extended MIR emission in these AGN. The methods can be applied to other MIRIM datasets which are similarly limited by dynamic range due to one or more very bright point-like components.

From the results presented in this work, we arrive at the following conclusions:

\begin{itemize}
	\item Our images demonstrate the diversity of emitting structures in these AGN, including nuclear point sources, large-scale dusty disks, compact star-forming knots, and AGN-influenced regions.
	
	\item All the AGN in our sample show clear, bright nuclear point sources irrespective of their optical spectral type. The SEDs of the point sources are consistent with the general predictions from models of pc-scale dust tori illuminated by a radiatively-efficient accretion disk. 
	
	\item In most of the AGN In our sample, we find extended emission aligned with the axis of the nuclear illumination and cospatial with the narrow-line regions (NLRs). These ``polar'' emission structures are typically 200--300 pc in projected extent, and are 2--10 times higher in surface brightness compared to the surrounding star-forming disks.
	
	\item The MIR colours of the NLR-aligned emission are redder than the star-forming disks, implying a somewhat cooler dust temperature. 
	
	\item There is a high spatial correlation between the extended emission in the MIR, which is primarily from dust continuum, and highly ionised gas in the NLRs of these AGN. Complementary studies based on our programme find evidence that the MIR dust emission is not purely heated by the central source, but may be substantially influenced by localised shock heating.
	
\end{itemize}

\section*{Acknowledgements}

We thank the anonymous referee for a constructive review that has improved the quality of this work. We thank Ariana Oprea for assistance in the final run of JWST data reduction.
DJR, HH, and SC acknowledge the support of the UK's Science and Technology Facilities Council (STFC) through grant ST/X001105/1. AAH and LHM acknowledge support from grant PID2021-124665NB-I00 funded by MCIN/AEI/10.13039/501100011033 and by ERDF A way of making Europe. SGB acknowledges support from the Spanish grant PID2022-138560NB-I00, funded by MCIN/AEI/10.13039/501100011033/FEDER, EU. AJB acknowledges funding from the “FirstGalaxies” Advanced Grant from the European Research Council (ERC) under the EU’s Horizon 2020 research and innovation program (Grant agreement No. 789056). E.K.S.H., C. P, L. Z., M. L. and D. D. acknowledge grant support from the Space Telescope Science Institute (ID: JWST-GO-02064). MPS acknowledges support from grants RYC2021-033094-I, CNS2023-145506, and PID2023-146667NB-I00 funded by MCIN/AEI/10.13039/501100011033 and the European Union NextGenerationEU/PRTR. CRA and AA acknowledge support from the Agencia Estatal de Investigaci\'on of the Ministerio de Ciencia, Innovaci\'on y Universidades (MCIU/AEI) under the grant ``Tracking active galactic nuclei feedback from parsec to kiloparsec scales'', with reference PID2022$-$141105NB$-$I00 and the European Regional Development Fund (ERDF). AA also acknowledges support from the European Union (WIDERA ExGal-Twin, GA 101158446). This work is based  on observations made with the NASA/ESA/CSA James Webb Space Telescope. The data were obtained from the Mikulski Archive for Space Telescopes at the Space Telescope Science Institute, which is operated by the Association of Universities for Research in Astronomy, Inc., under NASA contract NAS 5-03127 for JWST. Part of the work is based on observations made with the NASA/ESA Hubble Space Telescope, and obtained from the Hubble Legacy Archive, which is a collaboration between the Space Telescope Science Institute (STScI/NASA), the Space Telescope European Coordinating Facility (ST-ECF/ESAC/ESA) and the Canadian Astronomy Data Centre (CADC/NRC/CSA). ALMA is a partnership of ESO (representing its member states), NSF (USA) and NINS (Japan), together with NRC (Canada), NSTC and ASIAA (Taiwan), and KASI (Republic of Korea), in cooperation with the Republic of Chile. The Joint ALMA Observatory is operated by ESO, AUI/NRAO and NAOJ."

\section*{Data Availability}

All primary \jwst\ and {\it HST} data used in this work is publicly available from the Mikulski Archive for Space Telscopes (MAST) at  at \href{https://doi.org/10.17909/yv20-cm46}{doi.org/10.17909/yv20-cm46}. JWST science-ready images of all targets in all filter bands are available for download on the Newcastle University Research Data repository from the following URL: \href{https://doi.org/10.25405/data.ncl.32608851}{doi.org/10.25405/data.ncl.32608851}. {\it Spitzer}/IRS and VLT/SINFONI spectroscopic data are publicly available through the third-party databases referenced in this work. Primary science-ready ALMA data is available from the ALMA science archive. 

\section*{Affiliations}
$^{1}$ School of Mathematics, Statistics and Physics, Newcastle University, Newcastle upon Tyne, NE1 7RU, UK\\
$^{2}$ Department of Physics and Astronomy, University of Texas at San Antonio, One UTSA Circle, San Antonio, TX 78249, USA\\
$^{3}$ National Astronomical Observatory of Japan, National Institutes of Natural Sciences (NINS), 2-21-1 Osawa, Mitaka, Tokyo 181-8588, Japan\\
$^{4}$ Space Telescope Science Institute, 3700 San Martin Drive, Baltimore, MD 21218, USA\\
$^{5}$ Centro de Astrobiolog{\'i}a (CAB) CSIC-INTA, Camino Bajo del Castillo s/n, 28692 Villanueva de la Ca{\~n}ada, Madrid, Spain \\
$^{6}$ Max Planck Institute for Extraterrestrial Physics (MPE), Giessenbachstr.\ 1, 85748 Garching, Germany\\
$^{7}$ Department of Physics and Astronomy, University of Alaska Anchorage, Anchorage, AK 99508-4664, USA\\
$^{8}$ Observatorio Astron\'omico Nacional (OAN\mbox{--}IGN), Observatorio de Madrid, Alfonso XII 3, 28014 Madrid, Spain\\
$^{9}$ Department of Physics \& Astronomy, University of Southampton, Highfield, Southampton SO17 1BJ, UK\\
$^{10}$ Department of Physics \& Astronomy, University of South Carolina, Columbia, SC 29208, USA\\
$^{11}$ Instituto de F\'{\i}sica Fundamental (IFF), CSIC, Calle Serrano 123, 28006 Madrid, Spain\\
$^{12}$ Instituto de Astrof\'{\i}sica de Canarias, Calle V\'{\i}a L\'actea s/n, E\mbox{-}38205 La Laguna, Tenerife, Spain\\
$^{13}$ Departamento de Astrof\'{\i}sica, Universidad de La Laguna, E\mbox{-}38206 La Laguna, Tenerife, Spain\\
$^{14}$ Departamento de F\'{\i}sica de la Tierra y Astrof\'{\i}sica, Facultad de CC F\'{\i}sicas, Universidad Complutense de Madrid, E\mbox{-}28040 Madrid, Spain\\
$^{15}$ LUX, Observatoire de Paris, Coll\`ege de France, PSL University, CNRS, Sorbonne University, Paris, France\\
$^{16}$ Department of Physics, University of Oxford, Denys Wilkinson Building, Keble Road, Oxford OX1 3RH, UK \\
$^{17}$ Institute of Astrophysics, Foundation for Research and Technology Hellas (FORTH), Heraklion, 70013, Greece\\
$^{18}$ School of Sciences, European University Cyprus, Diogenes Street, Engomi, 1516, Nicosia, Cyprus\\
$^{19}$ Instituto de Radioastronom\'{\i}a y Astrof\'{\i}sica (IRyA), Universidad Nacional Aut\'onoma de M\'exico, Antigua Carretera a P\'atzcuaro \#8701, Colonia ExHda.\ San Jos\'e de la Huerta, Morelia, Michoac\'an, M\'exico C.P.\ 58089\\
$^{20}$ Telespazio UK for ESA, ESAC, Camino Bajo del Castillo s/n, 28692 Villanueva de la Ca\~nada, Spain\\
$^{21}$ Department of Astronomy, University of Geneva, ch. d'Ecogia 16, 1290, Versoix, Switzerland \\
$^{22}$ Departamento de F\'{\i}s\'{\i}ca, CCNE, Universidade Federal de Santa Maria, Av. Roraima 1000, 97105-900, Santa Maria, RS, Brazil\\
$^{23}$ LIRA, Observatoire de Paris, PSL Research University, CNRS, Sorbonne Universit\'e, Université Paris Cité, 5 place Jules Janssen, 92195, Meudon, France\\
$^{24}$ Astronomical Observatory, Volgina 7, 11060 Belgrade, Serbia\\
$^{25}$ Sterrenkundig Observatorium, Universiteit Gent, Krijgslaan 281-S9, Gent B-9000, Belgium\\



\bibliographystyle{mnras}
\bibliography{gatos_jwst_imaging_paper1}


\appendix

\section{{\it Spitzer}/IRS spectroscopy} \label{irs_spectra}

\begin{figure*}
	\centering
	\includegraphics[width=\textwidth]{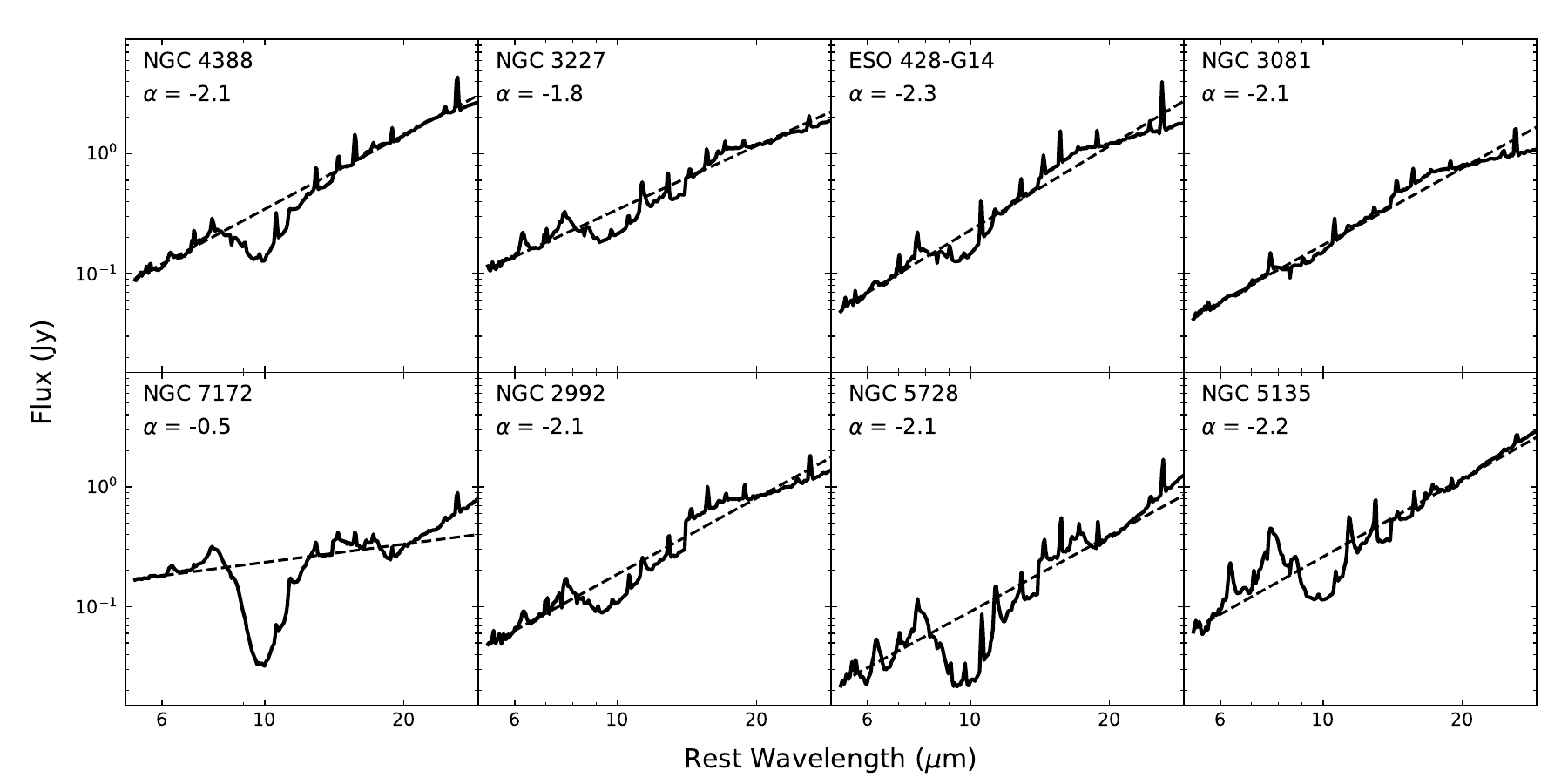}
	\caption{{\it Spitzer}/IRS low-resolution spectra of the eight Seyfert galaxies targetted by our JWST imaging programme. The dashed line demonstrates the equivalent power-law slope of the spectra evaluated from the average flux at 5.6 \mics\ and 21 \mics. The power-law slope is written in each panel. The spectra are presented in more detail in Figure \ref{nuclear_seds}. 
	}
	\label{IRS_spectra}
\end{figure*}

We downloaded {\it Spitzer} InfraRed Spectrograph (IRS) low-resolution MIR spectra of all our targets from the Combined Atlas of Source with Spitzer IRS Spectra (CASSIS) database, a value-added resource that hosts science-grade reductions of all spectra observed in the IRS staring mode. The {\it Spitzer} archive AORKEYs of all the spectra are given in Table \ref{irs_data}. Within CASSIS, these can be used to identify the individual reduced spectra.

For our analysis, we optimally-extracted one-dimensional spectra for a point-source with the {\it Spitzer} PSF. These spectra sample a large angular region at longer wavelengths than at shorter wavelengths, with extraction window sizes of $\approx 3$--$9$ arcseconds. 

We use the optimally-extracted low-resolution IRS spectra to validate the flux calibration of our \jwst\ images (Section \ref{nuclear_seds}) and to decide on a representative power-law spectral shape for the nuclear emission to use in our PSF modelling (Section \ref{psfs}). For this latter analysis, we calculate the slope of the power-law using the average IRS spectral flux at 5--6 \mics\ and 19--23 \mics, the approximate widths of the F560W and F2100W filters. The resultant power-laws are shown, along with the IRS spectra, in Figure \ref{IRS_spectra}. As one may notice, the power-laws capture the basic shape of the spectra across the MIR range of our MIRI imaging, despite the small jumps in flux that are found between some of the IRS spectral segments (a consequence of the extraction approach).

In some of our targets, the prominent silicate absorption feature influences the shape of the spectra around 10 \mics\, sampled by the MIRI F1000W filter. Detailed tests of the influence of the silicate feature on the resultant F1000W PSFs show that the effects are at worst a few percent of the unresolved nuclear flux in this band \citep{charidis25}. Therefore, for this work, we just adopt a power-law input spectrum for the modelling with an index as shown in Figure \ref{IRS_spectra}. 

\begin{table}
	\caption{Observation IDs for the {\it Spitzer}/IRS datasets}
	\centering
	\begin{tabular}{ccc}
		\hline
		Galaxy Name & AORKEY \\
		\hline
		NGC\,4388 & 18510848 \\
		NGC\,3227 & 4934656 \\
		ESO 428-G14 & 18507264 \\
		NGC\,3081 & 18509824 \\
		NGC\,7172 & 18513920 \\	
		NGC\,2992 & 4934144 \\
		NGC\,5728 & 18945536 \\				
		NGC\,5135 & 18512384 \\
		\hline			
	\end{tabular}
	\label{irs_data}
\end{table}

\section{Early JWST observations} \label{firstobs}

The first data for PID 2064 were taken in early July 2022, among some of the very first science observations
with \jwst. Three targets -- NGC\,5728, NGC\,5135 and NGC\,7172 -- were observed during these early days. 
On preliminary examination of these data, it was clear that the pointing for the observations in the long-wavelength filters was incorrect, leading to a large, systematic offset of the nucleus 
towards the lower right corner of the detector (high X and low Y in the {\it ideal} coordinate system defined for the MIRI instrument). Especially in the F1800W and F2100W images, significant parts of the circumnuclear regions of these galaxies were lost from view because their centres were being placed too close to the edge of the subarray field. This issue created a problem for our science goals which relied on uniform imaging of the extended structure in the sources.

Following a report to the \jwst\ team at STScI, it was established that a issue in the telescope operations was the cause
of the shifts. The problem appeared specifically in dithered sequences with MIRIM filters that increased progressively towards
longer wavelengths in time. Following a Webb Operations Problem Report (WOPR 88500), the Space Telescope ScIence Institute (STScI) granted us repeat observations of the three targets, which were successfully executed later in Cycle 1. More details can be found on the programme information page maintained by STScI (\url{www.stsci.edu/jwst/science-execution/program-information?id=2064})

In this paper, we only present details of the final ``correct'' observations, but interested readers should keep in mind
that another set of somewhat flawed, though equally sensitive, observations exist for NGC\,5728, NGC\,5135 and NGC\,7172.

\bsp	
\label{lastpage}
\end{document}